\documentclass[useAMS,usenatbib]{mn2e}
\usepackage{graphicx}    
\usepackage{subfig}
\usepackage{amssymb}
 \usepackage{multirow} 
\DeclareGraphicsExtensions{.ps,.pdf,.png}
\makeatletter
\def\fps@figure{htbp}
\makeatother

\newcommand*{\Msun}{\ensuremath{\mathrm{M_\odot}}}%
\begin{document}
\title[Galaxy Evolution at Constant Number Density]{The Evolution of Galaxies at Constant Number Density: A Less Biased View of Star Formation, Quenching, and Structural Formation}
\author[Ownsworth et al.]{Jamie~R.~Ownsworth$^{1}$\thanks{E-mail: ppxjo1@nottingham.ac.uk}, Christopher~J.~Conselice$^1$, Carl J. Mundy$^{1}$, Alice Mortlock$^{1,2}$, \newauthor William G. Hartley$^{1,3}$, Kenneth Duncan$^{1,4}$, Omar Almaini$^{1}$ \\
$^{1}$University of Nottingham, School of Physics and Astronomy, Nottingham, NG7 2RD, U.K. \\
$^{2}$SUPA\thanks{Scottish Universities Physics Alliance}, Institute for Astronomy, University of Edinburgh, Royal Observatory, Edinburgh, EH9 3HJ, U.K. \\
$^{3}$Institute for Astronomy, ETH Zurich, Wolfgang-Pauli-Strasse 27, CH-8093 Zurich, Switzerland \\
$^{4}$Leiden Observatory, Leiden, The Netherlands}
\date{Accepted ??. Received ??; in original form ??}
\pagerange{\pageref{firstpage}--\pageref{lastpage}} \pubyear{2014}
\maketitle

\label{firstpage}
\begin{abstract}
Due to significant galaxy contamination and impurity in stellar mass selected
samples (up to 95\% from $z=0-3$), we examine the star 
formation history, quenching time-scales, and structural evolution of 
galaxies using a constant number density selection with data from the UKIDSS
Ultra-Deep Survey field.     Using this 
methodology we investigate the evolution of galaxies at a variety of 
number densities from $z=0-3$.  We find that samples
chosen at number densities ranging from $3\times10^{-4}$ to 
10$^{-5}$ galaxies Mpc$^{-3}$ (corresponding to $z\sim0.5$ stellar masses of 
M$_{*}= 10^{10.95-11.6}$ M$_{0}$) have a star forming blue
fraction of $\sim50$\% at $z\sim2.5$, which evolves to a nearly $100$\% 
quenched red and dead population by $z\sim 1$.   We also see evidence 
for number density downsizing, such that the galaxies selected at the 
lowest densities (highest masses) become a homogeneous red population before
those at higher number densities.    Examining the evolution of the 
colours for these systems furthermore shows that the formation redshift of
galaxies selected at these number densities is $z_{\rm form}>3$.
The structural evolution through size and S\'{e}rsic index fits reveal that
while there remains evolution in terms of galaxies becoming larger and 
more concentrated in stellar mass at lower redshifts, the magnitude of 
the change is 
significantly smaller than for a mass selected sample. We also
find that changes in size and structure continues at 
$z < 1$, and is coupled strongly to passivity evolution.  We conclude that
galaxy structure is driving the quenching of galaxies, such that galaxies
become concentrated before they become passive.


\end{abstract}


\section{Introduction}
\label{sec:Intro}

In the local Universe, the most massive galaxies ($M_*>10^{11}\Msun$) are a nearly homogeneous population. They have early-type morphologies, red rest-frame optical colours, and low star formation rates (\citealt{Bower1992}, \citealt{Kauffmann2003}, \citealt{Gallazzi2005}, \citealt{Baldry2006}, \citealt{Conselice2006}, \citealt{Grutzbauch2011}, \citealt{O2012}, \citealt{Mortlock2013}).  A major unanswered question is: how have these massive galaxies evolved over cosmic time to become this population? 

 Recent measurements of the stellar mass function of galaxies out to $z=4$ (e.g. \citealt{Muzzin2013}, Ilbert et al. 2013, Mortlock et al. 2015, \citealt{Duncan2014}) show evidence that significant numbers of massive galaxies exist at very early cosmic times.  However, the total number densities of these massive galaxies grows substantially at later times, showing a drawn out formation history.    By redshift $z \sim 1$ the number densities of massive galaxies with M$_{*} > 10^{11}$ M$_{0}$ are consistent with their $z = 0$ values, demonstrating a rapid formation within the first half of the universe's history (e.g., Conselice et al. 2007; Mortlock et al. 2011; Mortlock et al. 2015).   This suggests that massive galaxies form some portion of their mass very early in the universe, and then assemble the remainder of their mass very quickly.  

Although the stellar mass functions of galaxies provide a simple and direct way to measure the abundance of a population and its overall growth as a function of time, it does not tell us how individual galaxies have assembled and evolved. Ultimately, one major goal is connecting local massive ``red and dead" galaxies to their progenitors at early cosmic times to examine how they evolved and changed in terms of their properties.   However, connecting the same galaxies over cosmic time remains a significant problem that has yet to be fully resolved.

Over the last decade there has been extensive research into the evolution of massive galaxies. These studies have shown that the more massive a galaxy is today, the earlier its star formation and merging must have completed and subsided, and the earlier its morphology become spheroidal (e.g. Bundy et al. 2005, 2006; \citealt{Renzini2006}; Conselice et al. 2008; Mortlock et al. 2013). This is  called ``Galaxy Downsizing", in which the most massive galaxies appear to be in place and stop forming in an apparently anti-hierarchical manner.  At high redshift, massive galaxies also appear to be different from galaxies with the same stellar mass at lower redshifts.   The massive high-z population consists of galaxies with low S\'{e}rsic indices, small sizes, and  high star formation rates (e.g. Trujillo et al. 2007; Buitrago et al. 2008; \citealt{Daddi2007}; Conselice et al. 2007; \citealt{Buitrago2013}, \citealt{Mortlock2013}) compared with
galaxies at similar masses in the low-z universe.   However, these results are
nearly all based on selecting galaxies at a constant stellar mass limit at all
epochs, typically M$_{*} > 10^{11}$ M$_{0}$.  

However, Mundy et al. (2015) recently showed that using a stellar mass limit
such as M$_{*} > 10^{11}$ M$_{0}$ leads to a significant precursor bias, such
that the sample  selected at low redshifts ($z \sim 0.5$) is 95\% contaminated with 
galaxies which were not in the sample at $z \sim 3$.  This precursor bias 
gets even worse at higher redshifts.  Using  a constant number density 
selection reduces the precursor bias by a factor of at least 
10 over using a constant stellar mass cut selection (Mundy et al. 2015).   Mundy et al. (2015) find
that while there is some contamination and incompleteness at the 50\% level,
even when using a constant number density selection, the stellar mass and
star formation properties remains the same to within a factor of two.  This
is compared with the factors of $> 10$ differences when using a stellar mass 
cut to trace the same star formation and average/total stellar masses within a 
selection (Mundy et al. 2015).



In this paper we use constant number density selections to  examine the evolutionary paths that distant massive galaxy progenitors have traveled to become the nearly homogeneous population we see today. With number density selection methods we aim to answer the questions: Do massive galaxies form in extreme star formation episodes in the early universe? At what cosmic epoch to they stop forming stars?  How many galaxies evolve from the blue cloud to the red sequence? How has their structure changed from high redshift? 

Recent work has begun to investigate the evolution of the properties of massive galaxies using number density techniques (e.g. \citealt{Papovich2011}, \citealt{Patel2013}, Conselice et al. 2013, \citealt{Marchesini2014}, Ownsworth
et al. 2014, \citealt{Papovich2015}).\, \cite{Marchesini2014} showed using number density selections that the progenitors of ultra massive galaxies (selected with log $M_{*} >11.8$) appear to have red $U-V$ colours, but also host large amounts of star formation (sSFR$>10^{-10}\,{\rm yr}^{-1}$) at $z>3$. They however find that the progenitors of ultra-massive galaxies, including the star forming objects, have never lived on the blue star-forming cloud in the last $\sim 11$ Gyr of cosmic history.  Papovich et al. (2011)
trace the star formation history of a luminosity based number density selection
from high-z to low.  In terms of galaxy formation, Conselice et al. (2013) investigate the gas accretion rate of a number density selected sample, and
Ownsworth et al. (2014) calculate the relative contribution of minor and
major mergers and gas accretion to galaxy formation within 
number density selected samples at $z < 3$.

Despite the importance of this approach, a general study examining the evolution of the most basic processes of galaxy formation - star formation histories, quenching, and the assembly of structure through time has not yet been done.  In this paper we investigate the evolution of galaxies with cosmic time of the progenitors of local massive galaxies with number volume densities from  $n=3 \times 10^{-4}$ Mpc$^{-3}$, to $n=10^{-5}$ Mpc$^{-3}$ 
(corresponding to a mass limit of M$_{*} > 10^{10.95}$ M$_{0}$ at $z = 0.5$) 
from $z=3$ to $z = 0$.  We use data from the UKIDSS Ultra Deep Survey (UDS) to investigate this question, utilizing the DR8 release with a 5 $\sigma$ depth of $K = 24.6$ over 0.77 deg$^{2}$.   

We investigate the evolution of our sample's colours, stellar masses, star formation rates, passivity and structural parameters over the redshift range of $0.3<z<3.0$.   We furthermore discuss how these characteristics change as a function of the initial co-moving density.   This is a companion paper to Ownsworth et al. (2014) where we investigate the merging history of the same galaxies through their mass assembly.

Throughout this paper we assume the cosmology $\Omega_{M}=0.3$, $\Omega_{\lambda}=0.7$ and $H_0=70$ km s$^{-1}$ Mpc$^{-1}$. AB magnitudes and a Chabrier IMF are used throughout.  This paper is divided into the following sections.  \S 2 discusses the data we use to carry out this analysis, including how we measure redshifts and stellar masses, and how we carry out our galaxy selection, \S 3 describes the results of the paper, and \S 4 is the summary.

\section{Data and Analysis}
\label{sec:Data}

\subsection{The Ultra-Deep Survey (UDS) Field}
\label{sec:UDS}

This work is based on the 8th data release (DR8) of the Ultra Deep Survey (UDS; Almaini et al in prep.), which is the deepest of the UKIRT (United Kingdom Infra\--Red Telescope) Infra\--Red Deep Sky Survey (UKIDSS; Lawrence et al. 2007) projects. The UDS covers 0.77 deg$^{2}$ in the J, H, K bands and the limiting magnitudes (AB), within an aperture of 2 arcsec at the 5$\sigma$ level, are 24.9, 24.2, 24.6 in J, H, K respectively. It is the deepest infra\--red survey ever undertaken over such a large area at these wavelengths. It benefits from an array of ancillary multi\--wavelength data: U\--band data from CFHT Megacam, B,V, R, i$^{\prime}$ and z$^{\prime}$ \--band data from the Subaru-XMM Deep Survey (SXDS; \citealt{Furusawa2008}); infrared data from the Spitzer Legacy Program (SpUDS). All of these are fundamental for the computation of accurate photometric redshifts, stellar masses and rest-frame magnitudes (e.g., Hartley et al. 2013; Mortlock et al. 2014). 

The galaxy catalogue employed in this work is K-band selected and contains approximately 96000 galaxies. As mentioned, this survey reaches a depth of K$_{AB}$=24.6 (5 $\sigma$ AB), which was determined from simulations and guarantees a 99\% completeness level (see \cite{Hartley2013} for more details).  The depth and wavelength of the UDS allows us to study the distant Universe with fewer biases against red and dusty galaxies, which could otherwise be completely missed in ultraviolet and optical surveys.

\subsection{Redshifts}
\label{sec:redshifts}

 The redshifts we use are a mixture of both photometric and spectroscopic redshifts.    The spectroscopic redshifts we utilise are from the UDSz redshift survey (e.g., Hartley et al. 2013; Almaini et al. 2016).  The spectroscopic redshifts from UDSz (ESO 180.A-0776) are measured through fitting templates to the spectra and using the best fit template for the redshift. We
however only use spectra  with a high certainty of having an accurate 
redshift measurement based on multiple emission lines or absorption features
in the rest-frame UV (Almaini et al. 2016, in prep). A further description of 
these redshifts is provided in Hartley et al. (2013).  In total there are
$\sim 1500$ spectroscopic redshifts in this programme. 

In addition to the redshifts from UDSz and other previous more focused programmes (see Hartley et al. 2015) we calculate photometric redshifts for our 
sample.   We use these photometric redshifts by fitting template spectra to 
photometry using \textsc{EAZY} \citep*{EAZY2008}. The template 
fitting we use is done with the standard six \textsc{eazy} templates and an extra blue one.  This blue template is a 
combination of the bluest \textsc{eazy} template and a small amount of
SMC-like extinction using A$_{V} = 0.1$ and the SMC Dust extinction law
\citep{Prevot1984}.   From inspection of UV spectra of $z > 2$ galaxies we
determine that this slightly altered template does a better job of fitting
some of these distant star forming galaxies.  We also used an interactive 
approach to determine the slight offsets in zero points for our photometric
bands to improve the agreements between the photometric and spectroscopic
redshifts.  Ultimately the photo-zs that we calculate from \textsc{EAZY} are
based on the maximum likelihood redshift after taking into account the K-band
apparent magnitude prior.  

We tested our photometric redshifts for quality in a few ways.  Our
basis for measuring the photometric redshift quality is through 
comparing with the $\sim$1500 
spectroscopic redshifts from UDSz and 
$\sim$4000 archival spectroscopic redshifts.  After we remove obvious AGN and catastrophic outliers ($\delta z/(1+z)>0.15$), we calculate that the dispersion between the photometric and the spectroscopic redshifts is $\delta z/(1+z)\sim0.031$ (\citealt{Hartley2013}).   Furthermore we performed the test of using
close pairs of galaxies as outlined in Quadri \& Williams (2010).  The idea
here is that galaxies which are close together on the sky are likely to be
at similar redshifts (see also Hartley et al. 2014).  We test our photo-zs
with this method and find a similar photo-z quality as given by the comparison
with spectroscopic redshifts.

\subsection{Stellar Masses \& SED fitting}
\label{sec:mass}

The stellar masses and rest\--frame colours (UVJ) of our sample are measured 
using a multi-colour stellar population fitting technique. For a full 
description see \cite{Mortlock2013} and \cite{Hartley2013}. We fit 
synthetic spectral energy distributions (SEDs) constructed from the stellar 
populations models of \cite{Bruzual2003} to the U, B, V, R, i$^{\prime}$, 
z$^{\prime}$, J, H, K  bands and IRAC Channels 1 and 2, assuming a 
Chabrier initial mass function. The star formation history is characterised 
by an exponentially declining model with various ages, metallicity and dust 
contents of the form

\begin{equation}
SFR(t=obs)= SFR_{\rm form} \times \rm{exp}(-t/\tau)  
\end{equation}

\noindent where the values of the fitted $\tau$ ranges between 0.01 and 
13.7 Gyr, and the age of the onset of star formation ranges from 0.001 to 
13.7 Gyr. We exclude templates that are older than the age of the Universe 
at the redshift of the galaxy being fit.   This declining star formation rate 
is well justified by previous
work showing that the observed star formation rate indeed declines 
exponentially at the galaxy co-moving densities we use in this paper 
(Ownsworth et al. 2014).

The metallicity used within the fitting ranges from 0.0001 to solar, and the dust content is parameterized, following \cite{Charlot2000}, by $\tau_{v}$, the effective V\--band optical depth. We use values up to $\tau_{v}=2.5$ with a constant inter\--stellar medium fraction of 0.3. 

We fit our SEDs by first scaling the template K-band apparent magnitude to the observed galaxy K-band apparent magnitude. We then fit each scaled model template in the grid of SEDs to the measured photometry of each individual galaxy. We  then calculate $\chi^{2}$ values for each template, and select the best fitting template, obtaining a corresponding stellar mass and rest\--frame luminosities. 

\cite{Hartley2013}, following the method from \cite{Pozzetti2010}, found the UDS $95\%$ mass completeness limit as a function of redshift given by: log $M_{\rm{lim}}=8.27+0.81z-0.07z^2$. Galaxies that fall below $M_{\rm{lim}}$ are not used in the subsequent analysis. 
We base our measured densities on the stellar mass functions from Mortlock et al. (2015) (Table 1). The resulting stellar mass limits for our study at the various number density selections are listed in Table 2.  Our selections are similar to the mass range used in the study of Papovich et al. (2015) who examine the properties of the progenitors of galaxies with MW and M31 masses.  Although we do not go as low mass as the Milky Way, the M31 mass in Papovich is log M$_{*} \sim 11$ at $z \sim 0$ and is thus just slightly lower mass than the systems recovered with our 3$\times 10^{-4}$ Mpc$^{-3}$ selection..

\subsection{Star Formation Rates and Dust Exinction}
\label{sec:SFRexplan}

The star formation rates used in this work are derived using the 
rest-frame UV luminosity. A full explanation of this technique can be 
found in \cite{Ownsworth2014}. We briefly explain the technique here.
The rest-frame UV light traces the presence of young and short-lived stellar 
populations produced by recent star formation. The star formation rates are
calculated from scaling factors applied to the luminosities. These 
scaling factors are dependent on the assumed IMF (\citealt{Kennicutt1983}). 
However, UV light is very susceptible to dust extinction and a careful dust 
correction has to be applied. The correction we use here is based on the 
rest frame UV slope.

The raw 2800\AA\, NUV star formation rates ($SFR_{2800,\rm{SED}}$) used in 
this paper are obtained from the rest\--frame near UV luminosities measured 
from the best fit SED model found in the stellar mass fitting. We determine 
the dust\--uncorrected SFRs, $SFR_{2800,\rm{SED,uncorr}}$, for 
$z = 0.5\--3$ galaxies from applying the \emph{Galaxy Evolution Explorer} 
(GALAX) NUV filter to the best fit individual galaxy SED.

To measure the SFR  we first derive the UV luminosity of the galaxies in our 
sample, then use the \cite{Kennicutt1998} conversion from 2800\AA\, 
luminosity to SFR assuming a Chabrier IMF:

\begin{equation}
SFR_{UV} (\mathrm{M_{\odot}yr^{-1}}) = 8.24 \times 10^{-29} L_{2800} (\mathrm{ergs\,\, s^{-1}\, Hz^{-1})} 
\end{equation}

\noindent To obtain reliable star formation rates in the rest-frame ultraviolet, we need to account for the obscuration due to dust along the line of sight.   The way we do this is largely outlined in \cite{Meurer1999} who found a correlation between attenuation due to dust and the rest-frame UV slope, $\beta$, for a sample of local starburst galaxies, such that

\begin{equation}
f_{\lambda} \sim \lambda^{\beta}
\end{equation}

\noindent where $f_{\lambda}$ is the flux density per wavelength interval and $\lambda$ is the central rest wavelength. Using the ten UV windows defined by 
Calzetti et al. (1994) we measure $\beta$ values from the best fitting SED template for each galaxy.   We can do this as the redshift range we are examining has well calibrated UV SED fits due to many of the input photometric bands lying in the UV part of the spectrum.   These $\beta$ values are then converted into a UV dust correction using the \cite{Fischera2005} (FD05) dust model. This dust calculation originates from the same method that we use to calculate the stellar masses, as a result our dust values are quantized into the the units in which we apply the dust extinction to our model SEDs.

Whilst we show and discuss the A$_{2800}$ dust extinctions in this paper, these can be converted using the relation between the extinction at other wavelengths. To ease comparison with other papers, the conversion of $A_{\rm V} = 0.49 \times A_{2800}$ is applicable.  More details of this method are discussed and presented in Ownsworth et al. (2014)
for the sample we use throughout this paper.

%


%
%
%

\subsection{Galaxy Structural Parameters}

We calculate structural parameters measured on ground based UDS K\--band images using \textsc{galapagos} (Galaxy Analysis over Large Area: Parameter Assessment by \textsc{galfit}ing Objects from SE\textsc{xtractor}; \citealt{Barden2012}). This program uses SE\textsc{xtractor} and \textsc{galfit} to fit S\'{e}rsic light profiles (\citealt{sersic1968}) to objects in the UDS field. The S\'{e}rsic light profile is given by the following equation:

\begin{equation}
\Sigma(R)=\Sigma_{e}\times \rm{exp}\left(-b_{n}\left[\left(\frac{R}{R_{e}}\right)^{1/n}-1\right]\right)
\end{equation}

\noindent Where $\Sigma(R)$ is the surface brightness as a function of the radius, $R$; $\Sigma_{e}$ is the surface brightness at the effective radius, $R_{e}$; $n$ is the S\'{e}rsic index and $b_{n}$ is a function dependent on the S\'{e}rsic index. The sizes (effective radius) are calibrated with galaxy sizes derived from the UDS area from the Hubble Space Telescope (HST) Cosmic Assembly Near\--infrared Deep Extragalactic Legacy Survey (CANDELS) (\citealt{Grogin2011}, \citealt{Koekemoer2011}) by \cite{Wel2012}. For a full description of this method see \cite{Lani2013}, where it is shown that the ground based size measurements are reliable for galaxies with $K < 22$ in the UDS.  Some of the galaxies at our highest redshifts are fainter than this, and we do not use those small fraction when calculating the structural parameters.   We previously discussed these results and fits in Ownsworth et al. (2014) where we describe the size evolution of this sample.  

As the redshift range which we probe is quite large, $1 < z < 3$, the rest-frame wavelength range we probe with the K-band varies significantly.  This produces a  morphological k-correction, whereby we are probing rest-frame J-band at $z \sim 1$ and the rest-frame V-band at $z \sim 3$.  To address this we also measure the morphological parameters in shorter bands - J and H, and find essentially the same structural parameters. This is consistent with previous results which show that the structure and morphology is very similar for galaxies redward of the Balmer break for both nearby and distant galaxies (e.g., Taylor-Mager et al. 2007; Conselice et al. 2011).

\subsection{Constant Galaxy Number Density Selection }

We define our galaxy sample in the same way as in \cite{Ownsworth2014} using a constant galaxy number density selection at redshifts $z < 3$. In principle, selecting galaxies at a constant number density directly tracks the progenitors and descendants of massive galaxies at all redshifts. Studies such as Leja et al. (2013) and Mundy et al. (2015) have shown that this technique is robust at recovering the properties of the progenitors of local massive galaxies when using semi\--analytic models.  These models trace individual galaxies evolving over the last eleven billion years. 

However, as shown in Mundy et al. (2015) when examining low and high redshift galaxies the selected systems only have an overlap of at most  50\%.   This means there is a 50\% contamination rate of galaxies that were not in the sample at high redshift but entered it at lower redshifts.  However,  the properties of the galaxies replacing initial members are very similar to those being replaced (see Section 3).

In this study we select and compare galaxies at three constant co-moving number density values of $n=3\times10^{-4}\, \rm{Mpc^{-3}}$, $n=10^{-4}\, \rm{Mpc^{-3}}$, and $n=0.1\times10^{-4}\, \rm{Mpc^{-3}}$ at redshifts $0.3<z<3$ in six redshift bins. We chose these number densities as a trade\--off between having a robust number of galaxies in the analysis at each redshift, and retaining a mass complete sample at the highest redshifts. This number density range is comparable to number densities used in other similar studies (e.g. \citealt{Papovich2011}, \citealt{C2013}, \citealt{Ownsworth2014}, Papovich et al. 2015). 

\begin{figure}
 
\subfloat[]{\includegraphics[scale=0.4]{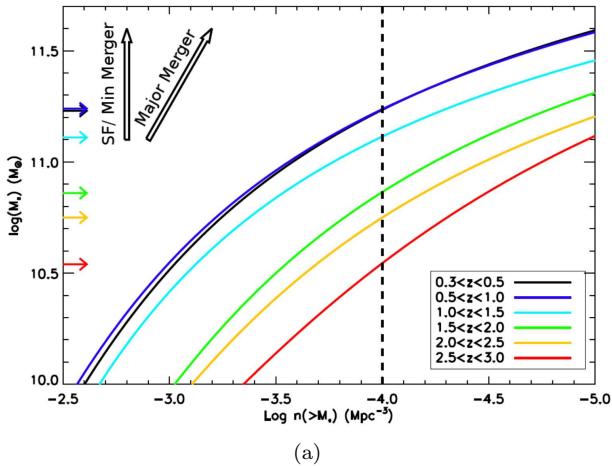}}
\caption{The integrated stellar mass functions from $z=0.3$ to $z=3$ from Mortlock et al. (2015). These integrated stellar mass functions gives us the co-moving number density of all galaxies more massive than at a given stellar mass. The large open black arrows indicate the expected evolution due to star formation, minor mergers and major mergers. We compare galaxies at a constant number density by selecting galaxies at each redshift at limits of $n(>M_{*}) = 10^{-4}\rm{Mpc^{-3}}$. The black dashed vertical line denotes the constant number density of $10^{-4}\rm{Mpc^{-3}}$. The coloured arrows indicate the values of $M_{*}$ that correspond to this number density for each integrated stellar mass fraction.}
\label{smf}
\end{figure}

We select our sample based on the integrated mass functions of the UDS field over the redshift range of $z= 0.3$ to 3.0 from Mortlock et al. (2014).   The stellar mass profile fits as a function of redshift in which we use to calculate the relationship between number density and mass is shown in Table~1.  Figure \ref{smf} shows the integrated mass functions from Mortlock et al. (2014) and the lower stellar mass limits for the constant number density selection. The values for the limits are also listed in Table~2. The arrows in the top left hand of Figure \ref{smf} show how the galaxy stellar mass functions will change due to stellar mass growth through star formation and 
merging.

\begin{center}
\begin{table}
\caption{Stellar mass function best fit Schechter function parameters from Mortlock et al (2015).}
\begin{center}
\begin{tabular}{| c | c | c | c |}
  \hline
  \hline
  $z$ & $\rm{log(M_{*})} \rm{(M_{\odot})}$ & log($\Phi^{*}$) & $\alpha$ \\
  \hline             
  $0.3-0.5$ & $11.32\pm0.07$ & $-3.20\pm0.08$ &$ -1.41\pm0.02$\\
  $0.5-1.0$ & $11.16\pm0.04$ & $-3.12\pm0.05$ & $-1.34\pm0.02$\\
  $1.0-1.5$ & $11.04\pm0.04$ & $-3.21\pm0.06$ & $-1.31\pm0.03$\\
  $1.5-2.0$ & $11.15\pm0.06$ & $-3.74\pm0.09$ & $-1.51\pm0.03$\\
  $2.0-2.5$ & $11.02\pm0.10$ & $-3.78\pm0.14$ & $-1.56\pm0.06$\\
  $2.5-3.0$ & $11.04\pm0.11$ & $-4.03\pm0.16$ &$ -1.69\pm0.06$\\
  \hline  
\end{tabular}
\label{tab:sfmevo}
\end{center}
\end{table}
\end{center}

\begin{center}
\begin{table}
\caption{Stellar mass limits for constant number densities used in this paper taken from the integrated mass functions shown in Figure \protect \ref{smf} .}
\begin{center}
\begin{tabular}{| c | c | c | c |}
  \hline
  \hline
  \multirow{2}{*}{Redshift ($z$)} & \multicolumn{3}{|c|}{Stellar Mass limit (log$\rm{M_{\odot}}$)} \\
   & $3\times 10^{-4}\rm{Mpc^{-3}}$ & $10^{-4}\rm{Mpc^{-3}}$ & $10^{-5}\rm{Mpc^{-3}}$\\
  \hline             
  $0.3-0.5$ & $10.95\pm$0.05 & $11.24\pm0.07$ & $11.59\pm$0.04\\
  $0.5-1.0$ & $10.97\pm$0.04 & $11.24\pm0.04$ & $11.58\pm$0.04\\
  $1.0-1.5$ & $10.84\pm$0.05 & $11.11\pm0.04$ & $11.45\pm$0.04\\
  $1.5-2.0$ & $10.51\pm$0.08 & $10.86\pm0.05$ & $11.31\pm$0.05\\
  $2.0-2.5$ & $10.40\pm$0.09 & $10.75\pm0.07$ & $11.20\pm$0.06\\
  $2.5-3.0$ & $10.16\pm$0.09 & $10.54\pm0.09$ & $11.11\pm$0.04\\
  \hline  
\end{tabular}
\label{tab:smfc}
\end{center}
\end{table}
\end{center}

\section{Results}

Using our galaxy selection methods based on the different values of the 
co-moving number 
density we examine the properties of galaxies selected through this
method.  That is, we examine the colour evolution, the passive galaxy 
fraction evolution, and the dust evolution between $z = 0.5 - 3$.

Before we discuss the properties of these galaxies, we give a brief background
to the analysis here and how our results can be interpreted.  First, Ownsworth
et al. (2014) studied the evolution of the stellar mass and star formation
rates for galaxies selected with a variety of number densities. As our
canonical co-moving number density we use in this paper is 
$n=10^{-4}\rm{Mpc^{-3}}$, we discuss briefly the results of Ownsworth et al.
(2014) where the mass evolution of this sample is examined. 
Other number densities give slightly different results, however, 
as outlined in the Appendix.  

Using a number density selection of $n=10^{-4}\rm{Mpc^{-3}}$ we find that the
mean stellar mass for this selection  changes from log M$_{*} = 10.6$ at
$z \sim 3$
up to log M$_{*} = 11.3$ at $z = 0$.  Over this time period
the mass of these galaxies grows by a factor of $\sim 4$.  This implies
that of the total stellar mass in a $n=10^{-4}\rm{Mpc^{-3}}$ selected sample
at $z = 0.5$, only 25\% of that mass would have been already 
within the galaxy at $z = 3$.  We show that a significant fraction 
of the mass in these galaxies
formed through other methods beyond
star formation, with the most obvious possibility being merging.  Ownsworth
et al. (2014) furthermore discuss how the stellar mass built up comes from
equal amounts of merging and gas accretion.  It is therefore now worth asking
the follow up question about the state of these galaxies as they evolve
through this time.

In a similar study, Mundy et al. (2015) investigate the reliability of
using a co-moving volume sample to examine the evolution of galaxies.  The
goal in Mundy et al. (2015) was to determine the fraction of galaxies selected
in a sample which remain in that sample at lower redshifts (purity)
and the contamination of new galaxies when using a number density selection.  
Mundy et al. (2015) find that it is impossible to retrieve exactly the
same galaxies through
cosmic time, with purity and completeness levels at $\sim 50$\% from $z = 3$
to $z = 0$.    However, when using a stellar mass cut, such as log M$_{*} =
11$ through all redshifts, the contamination fraction becomes as high as
95\% as early as $z \sim 1$ starting with a sample at $z \sim 3$ (Mundy
et al. 2015).  This means
that when using a constant mass cut to define a sample that only 5\% of the
galaxies at $z \sim 1$ are the descendents of the galaxies chosen with the
same selection at $z \sim 3$.  

However,  Mundy et al. (2015) showed that while a galaxy sample selected at a
constant number density selection can be contaminated, the properties of the
galaxies replacing the galaxies removed are very similar to each other.
 Mundy et al. (2015) show that the
average and integrated masses and star formation rates chosen through 
a number density selection is very close to values of the initially 
selected sample, to within 50\%, and often much lower (Mundy et al. 2015).
 This shows that while the samples are not the same through time, the 
properties inferred are similar to what would be measured if the identical
samples could be retrieved completely.  We therefore adopt this
approach of using number density selection for understanding the evolution 
of a galaxy population, with these caveats and assumptions spelled out.

\subsection{Colour Evolution}

\begin{figure*}
\includegraphics[scale=0.8]{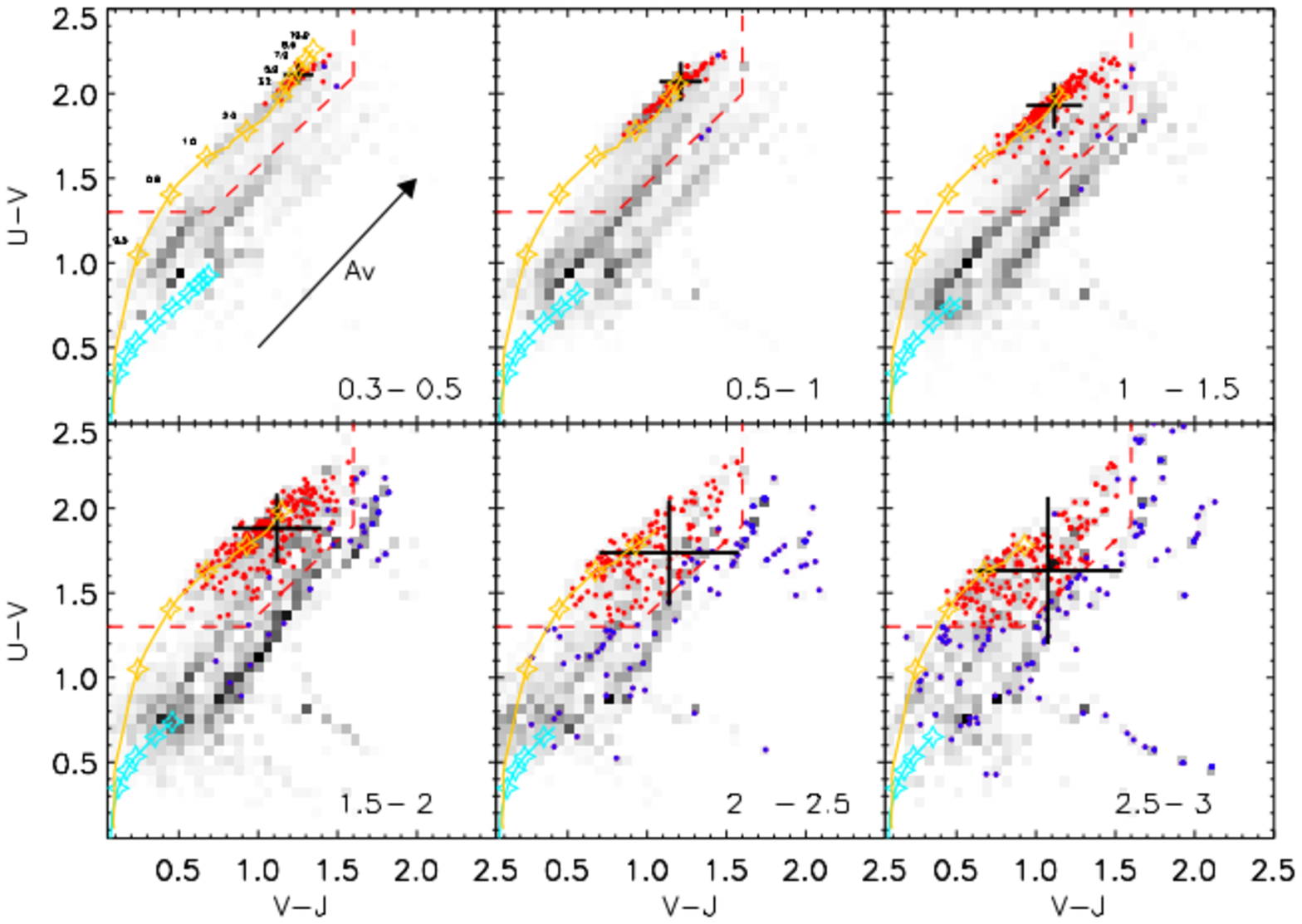}           
\caption[Rest frame $U-V$ versus $V-J$ diagram in redshift bins between $z=0.3$ and $z=3.0$ of the C\--GaND selected sample]{Rest frame $U-V$ versus $V-J$ diagram in redshift bins between $z=0.3$ and $z=3.0$ of the constant number density selected sample with $n=10^{-4}\rm{Mpc^{-3}}$.  This corresponding to a mass limit of log M$_{*} \sim  11.24$ at $z \sim 0.3$.   The red dashed lines denotes the $UVJ$ passive selection. Red circles show the progenitors of massive galaxies that are selected as passive via the $UVJ$ method. Blue circles show the progenitors of massive galaxies that are selected as star forming via the $UVJ$ method and $24\mu m $ criteria. The black cross shows the median colour and standard deviation for the progenitor sample in each redshift bin. Greyscale shows total population selected above the 95\% completeness limit within each redshift bin. The colour evolution tracks from \protect \cite{Bruzual2003} SSP models are also shown. The light blue line shows a constant star formation history with no dust and the yellow line shows an exponentially declining star formation history with $\tau=0.1\,\rm{Gyr}$. The blue open stars represent model colours at the specified ages, given in Gyr.  These have the same intervals as the orange line and are at: 0.5, 0.8, 1.0, 2.0, 3.5, 5.0, 7.0, 8.5, 10.0 Gyr. The colour evolution tracks are plotted up to the  age of the Universe in each redshift bin. Similar plots at other number densities can be found in the Appendix.}
\label{UVJ} 
\end{figure*}

\begin{figure*} 
\includegraphics[scale=0.8]{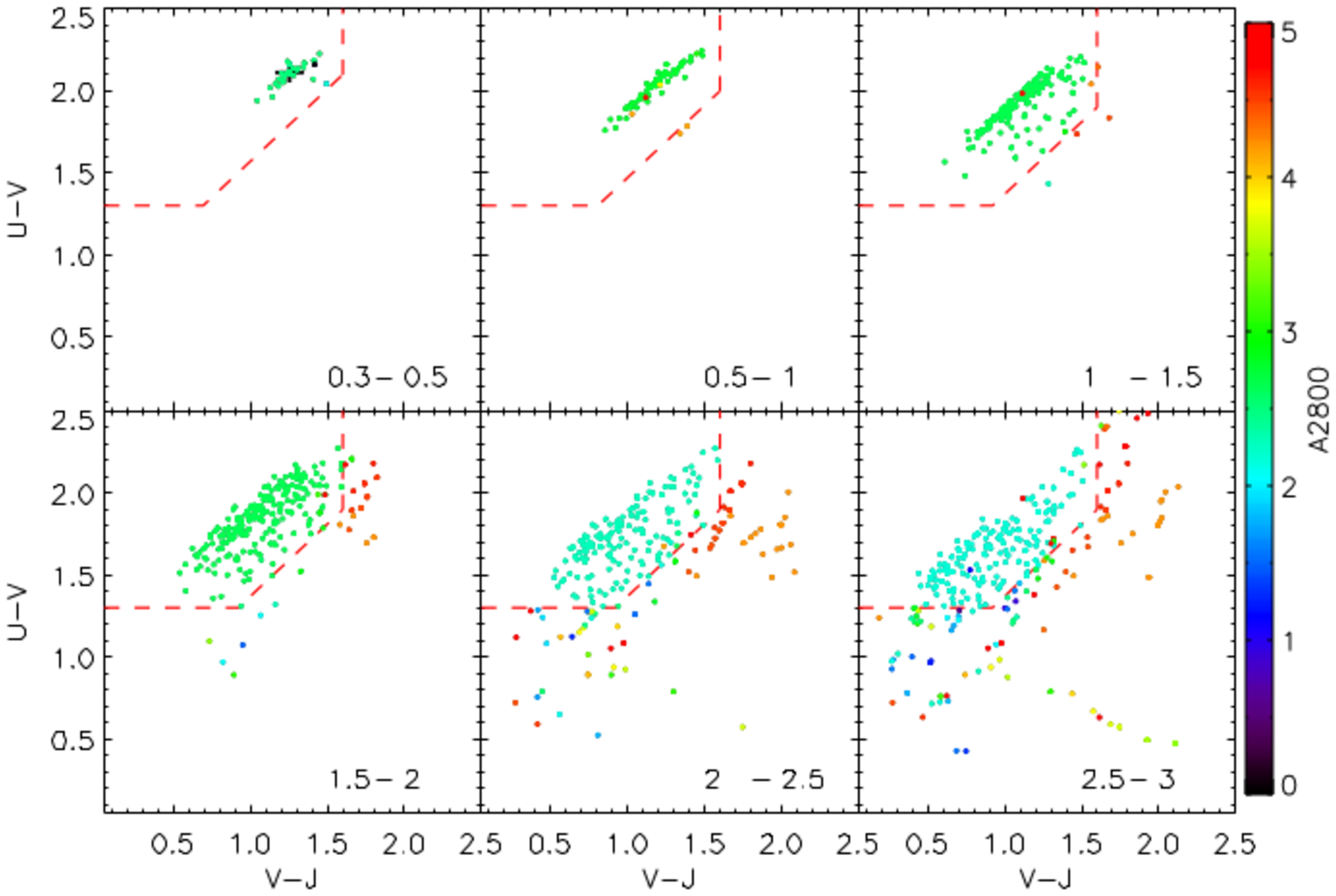}           
\caption[Rest frame $U-V$ versus $V-J$ diagram in redshift bins colour coded by $\rm{A_{2800}}$]{Similar to Figure \ref{UVJ}, but showing the dust content of each galaxy.  Rest frame $U-V$ versus $V-J$ diagram in redshift bins between $z=0.3$ and $z=3.0$ of the constant number density selected sample with $n=10^{-4}\rm{Mpc^{-3}}$. Coloured circles show the progenitors of massive galaxies with the colour representing the UV dust attenuation at 2800 \AA\, as shown by the colour bar on the right hand side.}
\label{UVJdust} 
 \end{figure*}

\subsubsection{Method}

The first thing we examine within our constant co-moving number density 
selection is the stellar populations of the galaxies selected through
this methodology.   

We do this in several ways, but the initial methodology for investigating
these galaxy's is through their position in the $UVJ$ digram.  
The rest-frame $U-V$ vs $V-J$ diagram is a useful tool to separate 
quiescent and star forming galaxies. It has become commonly used due 
to its ability to distinguish between truly quiescent objects and dust 
reddened systems at redshifts $z \sim 2$ (e.g. \citealt{Williams2009}).
Many alternative 
methods exist to separate a galaxy population into star forming and 
passive objects using broadband photometry e.g. g-r colour 
(\citealt{Bell2003}), u-r colour (\citealt{Baldry2004}), U-B colour 
(\citealt{Peng2010}) and $BzK$ colours (\citealt{Daddi2004}) see 
\cite{Taylor2014} for a comparison of these techniques.  We use the {\em UVJ}
method as it is has been used extensively at high redshifts and is therefore
the best understood in principle (e.g., Mortlock et al. 2015).  More details
in how we use the $UVJ$ diagram and its limitations is described for our
sample in Mortlock et al. (2015) and a further test is done in this paper
in \S 3.2.

The selection we use is based on the U, V and J Bessel band rest frame 
luminosities. These were also used by \cite{Williams2009} to select evolved 
stellar populations from 
those with recent star formation at $z < 2$. This technique is also used 
in \cite{Hartley2013} to extend the passive galaxy selection out to higher 
redshifts. The selection criteria for passive galaxies are as follows:

 \begin{equation}
U-V > 0.88 \times V-J+0.69	(z<0.5)
\end{equation}
\begin{equation}
U-V > 0.88 \times V-J+0.59	(0.5<z<1.0)
\end{equation}
\begin{equation}
U-V > 0.88 \times V-J+0.49	(z>1.0) 
\end{equation}

\noindent with $U-V >1.3$ and $V-J < 1.6$ in all cases. Although these 
criteria efficiently select galaxies with old stellar populations, there is a possibility that the `red' sample could still be contaminated by dusty star forming galaxies, edge on disks or AGN. We minimize this contamination by using the wealth of multi-wavelength data that is available in the UDS field. 

To identify active galaxies we cross match our sample with surveys taken 
at X-ray and radio wavelengths. For the X-ray we use data from the 
Subaru/XMM-Newton Deep Survey (\citealt{Ueda2008}) which covers the 
UDS field over the energy range of 0.5 keV to 10 keV. For the radio we 
use data from \cite{Simpson2006} which utilizes VLA 1.4 GHz data. 
We remove any galaxies that have either a detection in the X-ray or radio 
to clean this sample of AGN. This data will only effectively select out 
AGN at $z \lesssim 1$ due to the limits of these surveys, and will only be 
able to select the most radio loud and very active AGN at higher redshifts.

Furthermore the $24\rm{\mu m}$ data from the SpUDS provides a way to 
identify red objects that harbour dust-enshrouded star formation. 
Therefore any objects with a $24\rm{\mu m}$ detection 
($> 300\rm{\mu}$Jy, $15\sigma$) are assumed to be dusty star forming objects. 
Any galaxy found to be passive via the $UVJ$ selection criteria, but which has a bright $24\rm{\mu m}$ source associated with it will be reassigned to the star forming population and have a full UV dust correction applied.  We are
careful to exclude those objects that have an AGN signature, either through
X-ray emission or through signatures in those that have spectra. In 
total $\sim2\%$ of objects selected as being passive via the $UVJ$ criteria 
were reassigned to the star\--forming sample through this method.

\subsubsection{Stellar Population Ages}

Figure \ref{UVJ} shows the $UVJ$ diagram for the constant number density 
sample with $n=10^{-4}\rm{Mpc^{-3}}$ in different redshift bins. The 
red box region plotted in Figure \ref{UVJ} is from \cite{Williams2009} and 
denotes the passive galaxy selection.  Red points show galaxies that are 
selected as passive and blue points show galaxies that are selected as star 
forming within the given redshift bin. The large cross in each redshift 
plot denotes the median value for the whole progenitor population within 
each redshift bin. The greyscale shows the total population selected above 
the 95\% stellar mass completeness limit for the UDS sample 
from \cite{Hartley2013}.   

Our $UVJ$ diagram in Figure~\ref{UVJ} is similar to previous work
(e.g., Williams et al. 2009; Brammer et al. 2011; Marchesini et al. 2014; Papovich et al. 2015) with some exceptions. These previous studies in general are examining galaxies which contain a larger range, and thus lower, stellar masses than we examine in this paper.  If one restricts these previous diagrams to a high stellar mass limit, then one finds a good overlap as in Papovich et al. (2015) for the M31 progenitors, which are less massive than our nominal 
$n=10^{-4}\rm{Mpc^{-3}}$ selected sample, but more similar to our 
$n=3 \times 10^{-4}\rm{Mpc^{-3}}$ limit.  Furthermore, we only use this 
diagnostic to determine the difference between star forming and passive populations.

Although we do not discuss these results in this paper, our $UVJ$ colour
selection clearly correlates with galaxy morphology (passive=elliptical, star forming = disk+peculiar) (Margalef-Bentabol et al. 2016, submitted) 
as well as with galaxy clustering, whereby the passive
galaxies are clearly more clustered than the blue systems (e.g., Hartley et al. 2013; Wilkinson et al. 2016 submitted).  Thus this method does will in 
separating blue star formation from red passive systems.  

As can be seen from Figure~\ref{UVJ}, within the lowest redshift bin 
($z = 0.3-0.5$) the massive 
galaxy population constitutes a homogeneous population with extremely red 
$U-V$ colours with very little scatter. Moving to higher redshifts the 
scatter increases, and the population becomes more diverse in both $U-V$ 
and $V-J$ colours. However, as this population diversifies towards higher 
redshifts we find that the median $UVJ$ colour remains at all redshifts 
within the passive region, albeit with a larger scatter. When we compare
similar mass ranges to those of Papovich et al. (2015), which are presented in
the Appendix, we find a similar average colour evolution, to within or 
better than 0.3-0.5 dex,  in $(U-V)$
and $(V-J)$ colours, a difference which is within the uncertainties.   
However, Marchesini et al. (2014) find a 
significantly different pattern for the evolution of massive galaxies 
from this paper and Papovich et al. (2015). 


Also in Figure \ref{UVJ} we have plotted evolutionary tracks for 
the two colours from \cite{Bruzual2003} single stellar population models. 
The light blue line is a constant star formation history with no dust, 
and the yellow line is an exponentially declining star formation history with 
$\tau=0.1\rm\,{Gyr}$ and zero dust attenuation starting at the different
labeled look back times. Comparing with these models we find 
that at $z<0.5$ the progenitors of local massive galaxies harbour old 
(ages older than 5 Gyr) stellar populations which can be explained by an
exponentially declining star formation history. 

Examining the progenitors at higher redshifts, the median $UVJ$ colours 
within the error is always consistent with the exponentially declining models,
showing that a large fraction of this population is passively evolving. 
If we consider the effect of dust, the average age of the stellar populations 
for these galaxies
would decrease with increasing dust attenuation. As we move to higher 
redshift both the constant star formation evolution track with zero dust, and
the exponentially declining star formation history without dust does 
not accurately trace the whole star forming population, therefore this 
clearly indicates that the star forming progenitors must contain significant 
amounts of dust. 

In the Appendix we show the analog $UVJ$ diagrams for other number density 
selections, namely: $n=0.1\times10^{-4}\, \rm{Mpc^{-3}}$ and 
$n=3\times10^{-4}\, \rm{Mpc^{-3}}$. Both of 
these number density selections show similar behavior as the 
$n=10^{-4}\, \rm{Mpc^{-3}}$ selected sample. In the lowest redshift bin the galaxy 
population at all number density selections are a homogeneous 
population with red $U-V$ colours, with a small amount of scatter. 
Examining the $n=3\times10^{-4}\, \rm{Mpc^{-3}}$ and $n=0.1\times10^{-4}\, \rm{Mpc^{-3}}$ populations towards higher redshifts we find a similar 
result as the $n=10^{-4}\, \rm{Mpc^{-3}}$ galaxy population. 
The median $UVJ$ colour remains at all redshifts within the passive region.

\subsubsection{Dust Extinction}
z
In Figure \ref{UVJdust} we examine the dust extinction properties of the 
progenitor galaxy sample. In Figure \ref{UVJdust} the progenitor galaxies 
are colour coded to represent their dust extinction at $2800\,\rm{\AA}$ 
(A2800) measured from the UV slope. The uniformity of the passive objects in 
Figure \ref{UVJdust} arises from the method we used to derived the dust 
correction for these objects (e.g., see Ownsworth et al. 2014). 
Of the objects that 
are selected as star forming systems we find that at $z>1.5$ there is a diverse 
population of objects from dust poor objects lying towards the bottom left 
hand corner to highly dust attenuated systems lying towards the top right 
hand corner as expected for the $UVJ$ colour selection. The total star 
forming population at $z>1.5$ has an average $2800\rm \AA$ dust correction 
of $\sim 3.7$ mag.

We find a significant evolution in dust content over the redshift range $1.5<z<3.0$ 
for dust poor objects, those with a low $V-J$ colour.  These dust poor
objects are quite abundant at $z \sim 2.5$, with 
$28\pm4\%$ of star forming galaxies with $V-J < 1.0$, and decreasing 
towards $z=1.5$, where only $6\pm2\%$ of star forming galaxies have 
$V-J < 1.0$. We also find that a small population, $10\pm4\%$, of the 
star forming progenitors show rest-frame $U-V$ colours redder than, or
as red as, the 
quiescent progenitors.  

At higher redshifts, $z>2.5$, these objects span a wide range of 
rest frame colour values. Examining the derived UV-slopes
for the star forming population we find that the fraction of highly 
attenuated systems increases with higher redshift, similar to the result 
before. We find that $5\pm3\%$ of the star forming population at $z=3$ have 
$A_{2800}>5$ mag, increasing to $14\pm4$\% at $z=1.5$. This is 
accompanied by a decrease in the low dust attenuated systems, 
with $12\pm3\%$ of the star forming 
population with $A_{2800}<2$ mag at $z=3$, decreasing to $2\pm2\%$ at 
$z\sim 1.5$.  This suggests that the star forming progenitors at this 
redshift contain a wide range of dust and star formation properties 
unlike their low redshift descendants (see also \citealt{Whitaker2012b}, 
\citealt{Kaviraj2013}). We explore this in more detail in relation to the 
stellar mass of these systems later in this paper.

\subsection{Evolution in Colour vs. Stellar Mass}

\begin{figure*} 
\includegraphics[scale=0.8]{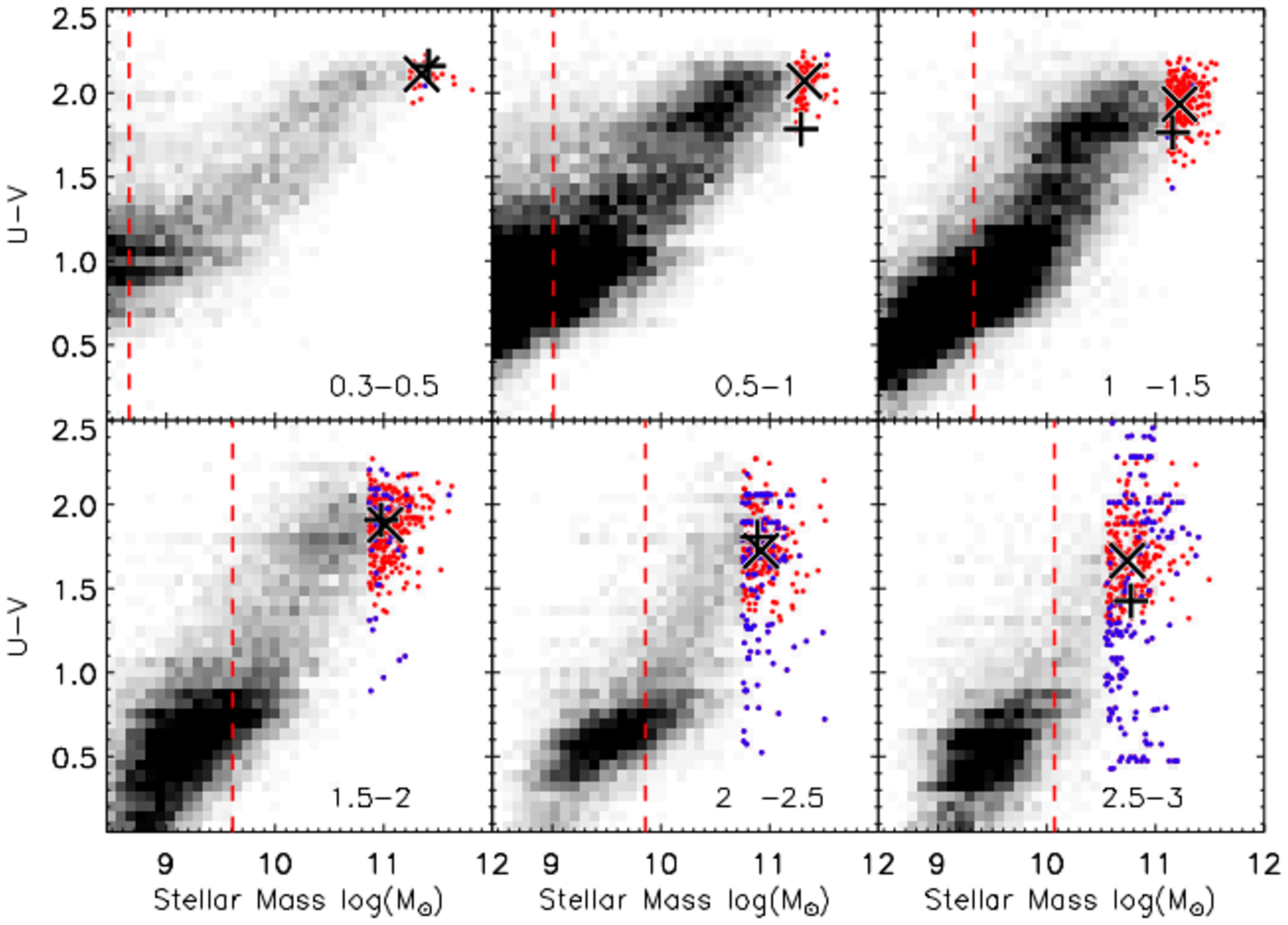}           
\caption[Stellar mass versus rest frame $U-V$ colour for all galaxies selected via the ]{Stellar mass versus rest frame $U-V$ colour for all galaxies selected via the constant number density selected sample with $n=1\times10^{-4}\rm{Mpc^{-3}}$. The red circles show the progenitors of massive galaxies that are selected as passive via the $UVJ$ method. The blue circles show the progenitors of massive galaxies that are selected as star forming via the $UVJ$ method. The black `X' shows the median $U-V$ colour for the passive population and the black plus `+' sign shows the median $U-V$ colour for the star forming population. The greyscale shows the whole UDS galaxy sample within each redshift bin. The red dashed line shows the 95\% stellar mass completeness limit. }
\label{sfrvsmass}
\end{figure*}

As highlighted in the previous section, the progenitors of local massive 
galaxies at relatively low redshift ($z < 1$) have similar colours, typical 
of quiescent and old 
stellar populations. As we look towards higher redshifts, some 
progenitors at our constant number density of $n=10^{-4}\rm{Mpc^{-3}}$
become star forming (\S 3.1). We find that some of the star forming 
progenitors exhibit a wide range of $U-V$ colours. We examine this result 
in a different way in Figure~4 using the $U-V$ rest frame 
colour versus stellar mass.  Figure \ref{sfrvsmass} shows the star forming 
and quiescent samples selected in the same way as in Figure \ref{UVJ}. The 
red dashed line shows the $95\%$ stellar mass completeness limit within each 
redshift interval. The blue points show the star forming progenitors with the 
median of this population represented by the black plus symbol `+'. The red 
points show the quiescent progenitors with the median of this population 
represented by the black cross, `X'. The greyscale show the total galaxy 
population within each redshift interval. 

We find that at the lowest redshift, the massive galaxy progenitors 
have very small scatter in both colour ($\sim0.08$ mag) and stellar mass, 
with the scatter increasing at higher redshifts.  In fact at this epoch
the mean colour and stellar mass for the blue and red systems is statistically
identical, although there are very few blue systems to compare with at the
lower redshifts.  

\begin{figure*} 
\vbox to 170mm{
\hspace{0cm} \includegraphics[angle=0, width=180mm]{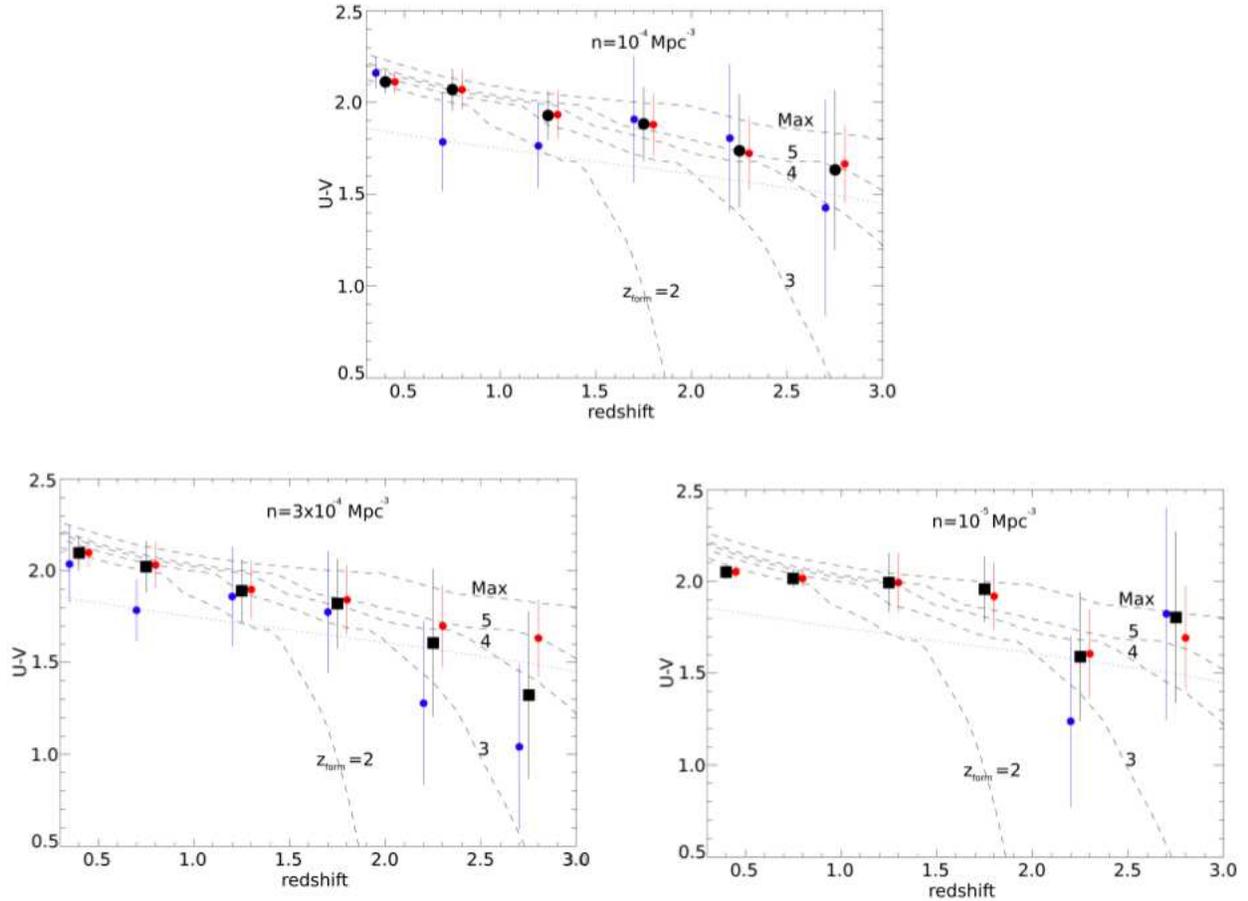}  \\  
\caption[Median r-frame $U-V$ colour versus redshift ]{Median rest-frame $U-V$ colour versus redshift for constant number density selected samples at three different
number densities. The black circles and squares show the evolution of the median $U-V$ colour of the whole progenitor population. The red and the blue circles show the evolution of the median $U-V$ colour of the passive and star forming samples respectively. Also shown is the colour evolution tracks from \protect \cite{Bruzual2003} models for a exponentially declining star formation
history starting at the labeled redshift. The black dashed lines show the colour evolution of a declining star formation history and varying formation redshifts starting from the beginning of the Universe (Max) to $\rm{Z_{form}}=2$. The light blue dotted line shows the colour evolution of a constant star formation history and $A_{\rm{v}}=2$ mag of dust extinction. This level of dust extinction is equivalent to the average dust correction of the star forming progenitors.}
}\label{UVz} 
 \end{figure*}

Examining this trend at higher redshifts the median for both the star forming 
and passive population do not show a large evolution, with the median 
$U-V$ colour of the star forming progenitors becoming bluer by 
$0.7\pm0.6$ mag over $0.3 < z < 3$, 
and the median colour for the passive progenitors becoming bluer by 
$0.5\pm0.2$ mag over the same epoch.  Figure \ref{sfrvsmass} demonstrates 
that the average star forming progenitor has a similar optical colour as a 
passive progenitor at the same redshift. Figure \ref{sfrvsmass} also shows 
that the average star forming progenitor has not lived in the blue star 
forming cloud since at least $z=3.0$, although there are a significant
number that do.

However, upon examining the population of star forming progenitors in more 
detail we find that $27\%$ at $z=3.0$ have blue, $U-V<1.0$, colours comparable 
to galaxies living on the $z=3.0$ blue cloud. Conversely, $24\%$ of the star 
forming progenitors at $z=3.0$ also have extreme red, $U-V>2.0$, colours. 
Examining the galaxies in our selection at $z=3.0$ that have red 
($U-V$) colours we find that the star forming systems are more numerous 
than the passive $UVJ$ selected progenitors by a ratio of $3:1$. 
The larger scatter in 
$U-V$ colours of the star forming progenitors is more pronounced than in the 
passive progenitors i.e. $0.6$ mag for star forming and $0.2$ mag for passive 
at $z=3.0$. The evolution in scatter between low and high redshift shows that 
the local red sequence is in the process of assembly between $0.3<z<3.0$.

In the Appendix we show the $U-V$ rest frame colour versus stellar mass of 
the $n=10^{-5}\, \rm{Mpc^{-3}}$ and 
$n=3\times10^{-4}\, \rm{Mpc^{-3}}$  selections. 
We find that the lower number density galaxy sample of 
$n=0.1\times10^{-4}\, \rm{Mpc^{-3}}$ has a very small scatter 
($\sim0.2$ mag )in $U-V$ colours across the whole redshift range studied. 
This suggests that these very high mass galaxies have undergone the majority 
of their colour evolution at $z>3$ (e.g., Duncan et al. 2014). 
The higher number density galaxy sample 
of  $n=3\times10^{-4}\, \rm{Mpc^{-3}}$, sampling typically lower mass galaxies, shows an increasingly large scatter towards higher redshifts with $31\%$ of the progenitors of local $n=3\times10^{-4}\, \rm{Mpc^{-3}}$ galaxies lying on the blue cloud with blue, $U-V<1.0$, colours. Compared to the $n=10^{-4}\, \rm{Mpc^{-3}}$ galaxy population the star forming progenitors of the $n=3\times10^{-4}\, \rm{Mpc^{-3}}$ galaxy population transition onto the red sequence at lower redshifts. This in an indication of galaxy ``downsizing" which can be seen within galaxy selection based on number density in addition to stellar mass.

In Figure~5 we show how the median $U-V$ colours for the total 
(black squares), star forming (blue circles) and, passive (red circles) 
evolve with redshift. Also plotted are the $U-V$ colour evolution tracks 
derived from \cite{Bruzual2003} SSP models with a exponentially declining 
star formation history as shown in Figure \ref{UVJ} plotted as the black 
dashed lines, and one with a constant star formation history with 
$A_{\rm{v}}=2$ mag of dust extinction, comparable to the average dust 
correction of the star forming population, shown by the light blue dotted 
line in Figure \ref{UVJ}.

These model tracks have a varying formation redshift from the 
beginning of the Universe (Max) down to $\rm{Z_{form}}=2$. The total 
population progenitors show a gradual evolution in their $U-V$ 
colours towards redder colours at lower redshifts, indicative of an 
aging stellar population that formed at redshifts of $z>4$. 
Dividing the population into star forming and  passive we find that 
the passive population follows the passively evolving colour tracks 
with hints that they may have stopped actively forming stars at redshifts as
high as $z=5$. 

We also examine the effects of increased dust extinction on these age
derivations.  The overall effect of dust is to decrease the 
formation redshift.   The average colours at low redshift are 
consistent with some
dust extinction, around A$_{\rm NUV} = 1.5$, as we found through our fits
to the SEDs of these galaxies.  When examining higher redshifts, where
there is more parameter space available for different scenarios, we find that
the colours are consistent with a $z_{\rm form} = 5$, but with no dust
extinction. As we do find some dust absorption through the SED fits,
 then a more realistic scenario also consistent with our colours 
is a formation redshift of $z_{\rm form} = 4$, with an extinction of
A$_{\rm NUV} = 1$, or a formation redshift of $z_{\rm form} = 3$ with 
an extinction of A$_{\rm NUV} = 2$.  Therefore these systems have a 
formation redshift of $z_{\rm form} > 3$.

The other number densities shown in Figure~5 show a similar pattern, but with
 the lower number densities at $n = 10^{-5}$~Mpc$^{-3}$ having a 
higher formation redshift than the 10$^{-4}$~Mpc$^{-3}$ selected systems, 
whilst the density selection of
$n = 3\times 10^{-4}$~Mpc$^{-3}$ has a lower redshift of formation than the
10$^{-4}$~Mpc$^{-3}$ selection. This is 
another indication of the downsizing, but seen here through number density
selections as opposed to stellar mass.

While the star forming population appears to be following 
the declining star formation history colour evolution tracks, they are also 
consistent with the dust reddened constant star formation history colour 
evolution track. However, from Figure \ref{UVJ} we see that they are not 
consistent with the exponentially declining star formation history
when examined in combination with other colours.

This result shows that a population selected at a constant number density
has formed the majority of its $z=3$ stellar mass
on average within the first Gyr 
of cosmic time. Is this plausible given our knowledge of the global cosmic 
star formation history? If we assume these objects formed their $z=3$ stellar 
masses over the redshift range $5<z<9$ ($\sim0.6$ Gyr) via star formation, 
the average SFR this implies is $114\,\rm{M_{\odot}yr^{-1}}$. Incorporating 
the number density of the progenitor galaxies, 
$n=10^{-4}\,\rm{Mpc^{-3}}$, gives a SFR density of these objects of 
$\rho_{\rm{SFR,progenitors}}=0.01\,\rm{M_{\odot}yr^{-1}Mpc^{-3}}$. From 
various works (e.g.  \citealt{McLure2013}, \citealt{Duncan2014}) the global 
cosmic SFR density over the redshift range $5<z<9$ varies from 
$\rho_{\rm{SFR,cosmic}}=0.05\pm0.03\,\rm{M_{\odot}yr^{-1}Mpc^{-3}}$ at 
$z=5$ to $\rho_{\rm{SFR,cosmic}}=0.02\pm0.06\,\rm{M_{\odot}yr^{-1}Mpc^{-3}}$ 
at $z=9$. As the global cosmic SFR density is larger than the SFR density 
inferred for the progenitor galaxies, it is therefore possible for these 
objects to form via star formation within the first Gyr of cosmic time.

\subsection{Star Formation History}

\begin{figure} 
\includegraphics[scale=0.4]{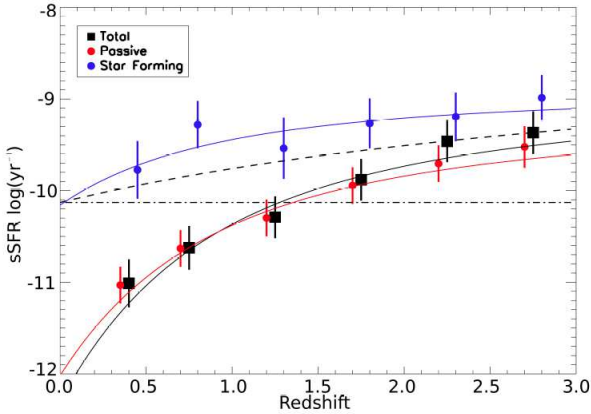}           
\caption[sSFR versus redshift for all galaxies selected via the 
C\--GaND]{The average specific star formation rate (sSFR) versus redshift 
for all galaxies selected via the constant number density selected sample with $n=10^{-4}\rm{Mpc^{-3}}$. Black squares show the evolution of the whole population. Red circles show galaxies that are selected as passive via the $UVJ$ method. Blue circles show galaxies that are selected as star forming via the $UVJ$ method. The horizontal dot-dashed line represents a stellar mass doubling time equal to the age of the universe at $z=0$. The dashed line represents a stellar mass doubling time equal to the age of the universe at a given redshift. The solid red, blue and black lines show the best fit exponentially declining star formation histories for the passive, star forming and total progenitor population respectively (see text). The errors of the fractions are derived from Monte Carlo analyses.}
\label{ssfr}  
 \end{figure}

\begin{figure} 
\includegraphics[scale=0.5]{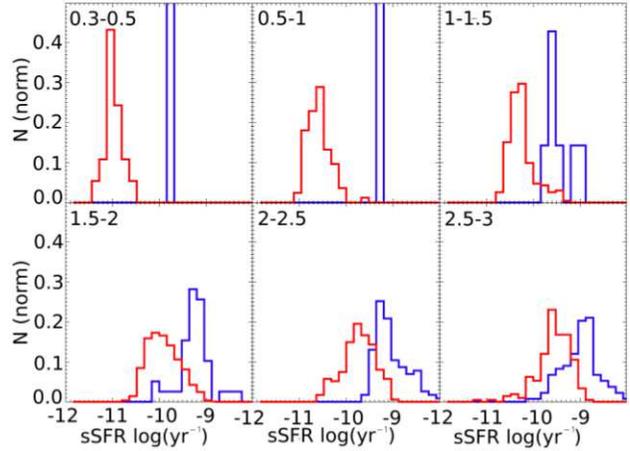}           
\caption[Histograms of the sSFRs of the $UVJ$ defined passive and star forming progenitor galaxies]{Histograms of the sSFRs of the $UVJ$ defined passive and star forming progenitor galaxies over the redshift range $0.3<z<3.0$ split into six redshift bins, which are labeled. The red histogram shows the sSFRs of the progenitors of local massive galaxies that are defined as passive via $UVJ$ colour selection and blue shows those that are classified as star forming. Both the passive and star forming histograms are normalised to the number of objects in each selection. }\label{ssfr2}  
 \end{figure}

\begin{figure} 
\includegraphics[scale=0.4]{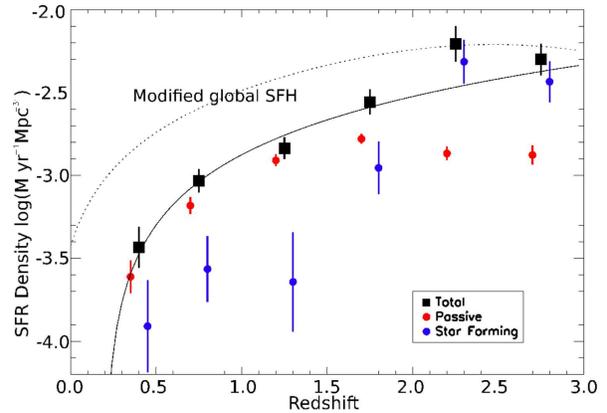}           
\caption[Star formation density versus redshift for all galaxies selected via the C\--GaND]{Star formation density versus redshift for all galaxies selected via the constant number density selected sample with $n=10^{-4}\rm{Mpc^{-3}}$. The black squares show the evolution of the whole galaxy sample and the red and blue circles show the evolution of the star formation density of the passive and star forming populations selected through $UVJ$ colours. The errors on the densities are derived from Monte Carlo analysis. The dotted line shows the global star formation history from \protect \cite{Hopkins2006} modified by $-1.5$ dex for clarity. The solid black line represents the best fit to the star formation density evolution of the total progenitor galaxy population.}

\label{sfrdens} 
 \end{figure}

Using our knowledge from the previous sections we now examine how and when 
the progenitors of local massive galaxies became the quiescent objects we 
see today. In this section we  examine the $n=10^{-4}\rm{Mpc^{-3}}$ 
number density sample.

Figure \ref{ssfr} shows how the average specific star formation rate 
(sSFR=$SFR/M_*$) of the total, star forming and, passive progenitor 
galaxies evolved from  $z=3.0$ for our sample using the number
density selection $n = 10^{-4}$ Mpc$^{-3}$. The blue circles show the 
median sSFR 
of the $UVJ$ selected star forming progenitor galaxies, the red circles 
show the median sSFR of the $UVJ$ selected passive progenitor galaxies and 
the black squares show how the median sSFR of the whole population evolves 
across this redshift range.   For the higher and lower number densities
discussed in the Appendix, we find essentially the same pattern.
Also shown in Figure \ref{ssfr} are lines 
denoting different stellar mass doubling times, i.e. the time it to takes 
for ongoing SFR to double the stellar mass of a given galaxy. The dot-dashed 
line denotes a doubling time equal to the age of the Universe at $z=0$, a 
passivity selection made in the local Universe. The dashed line shows a 
doubling time equal to the age of the Universe at a given redshift. 
Note that this doubling time at a given redshift appears to be a 
good dividing line between $UVJ$ passive and star forming systems. 

Not surprisingly, we find that the evolution of the sSFRs of the passive progenitor galaxies 
is faster than for star forming systems. The passive progenitor
 galaxies' median sSFR decreases with redshift by $1.5\pm0.3$ dex from 
$z=3.0$. The star forming progenitor galaxies median sSFR also decreases 
over the same time interval by only $0.8\pm0.4$ dex. If we examine the 
divide between the two populations, at low redshifts the difference in 
sSFR is more pronounced than at higher redshifts, with $\Delta \rm{sSFR} = 1.2\pm0.2$ dex at $z=0.3$ and $\Delta \rm{sSFR} = 0.5\pm0.4$ dex at $z=3.0$. 
We quantify the sSFR histories of the progenitor galaxies by fitting an 
exponentially  declining model of the form:

\begin{equation}
sSFR(t)=sSFR_0\times\rm{exp}(-t/\tau)
\end{equation} 

\noindent with $\tau=1.9\pm0.8$ Gyr for the total progenitor galaxy
population, $\tau=2.1\pm0.4$ Gyr for the passive objects and 
$\tau=4.7\pm0.5$ Gyr for the star forming objects. The larger value of 
$\tau$ for the star forming sample, compared to the passive objects, is 
as expected for a star forming population (Ownsworth et al. 2014). 

Using our knowledge of the sSFRs of the progenitor galaxies, in 
Figure~7 we examine the validity of the $UVJ$ colour selection. 
Figure~7 shows the normalised histograms of the passive and 
star forming populations as defined via the $UVJ$ colour selections 
across the redshift range we study.  We find that both populations appear
to be single peaked distributions across the redshift range studied, with 
an increasing overlap towards higher redshifts. Therefore, 
 the $UVJ$ colour selection appears to be an effective 
measure in separating the two populations for these massive galaxies
at the redshift ranges we study.  We find a similar result for the
other number densities we consider in the Appendix.

We also examine the evolution of the SFR density of these progenitors of 
massive galaxies. Figure \ref{sfrdens} shows the evolution of the SFR 
density with redshift. The black squares show the evolution of the total 
progenitor population and the red and blue circles show the passive and 
star forming objects respectively. Also shown in Figure \ref{sfrdens} is 
the global SFR history (SFH) from \cite{Hopkins2006} using the form from 
\cite{Cole2001}, $\rho(t)=(a+bz)h/(1+(z/c)^d)$ with $a=0.017$, $b=0.13$, 
$c=3.3$, $d=5.3$. The solid black line shows the best fit to the 
total progenitor population with the same form as the global SFH. We do 
not fit the SFR density evolution of the passive and star forming 
populations as their evolution is driven by their individual abundances as 
well as their star formation history. Therefore, the evolution of the 
passive and star forming SFR densities will not trace the same objects 
at all redshifts.  We find that the progenitors of local massive galaxies 
appear to undergo a sharper decrease in their SFR density than the global 
galaxy population SFH. They also show evidence that their SFH peaks at a 
higher redshift than the global galaxy population SFH. Both of theses 
findings are evidence for the downsizing scenario of galaxy formation.   

\subsection{Passive Fraction Evolution}
 
We examine in this section the passive fraction for our canonical  
$n=10^{-4}\rm{Mpc^{-3}}$ selection, as well as the higher and lower density 
selections.    We do this by using the information 
in the previous subsections including specific star formation rates, and
the passivity vs. star formation nature derived from the $UVJ$ colour 
selection.     In Figure \ref{pass} we show evolution of the $UVJ$ defined passive 
fraction of the progenitors of local massive galaxies.  The red boxes
show the fraction of galaxies that are selected as passive via this work. The 
red line is the best fit to the fraction with the form:

\begin{figure}
\includegraphics[scale=0.45]{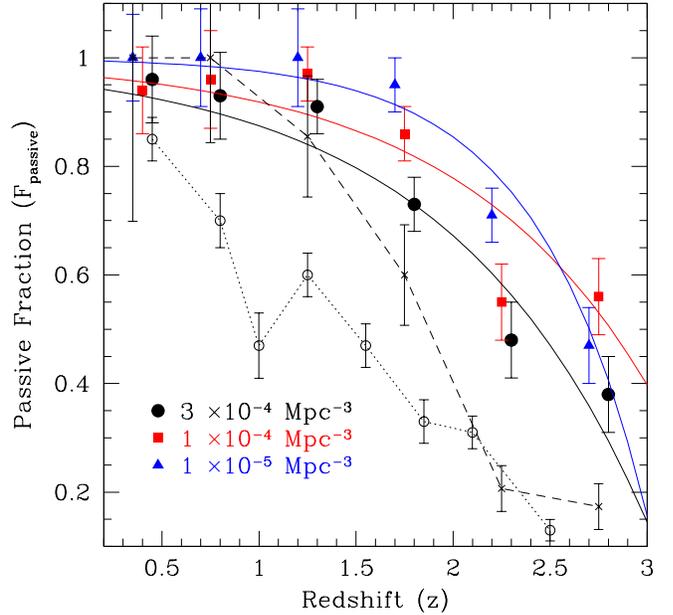}           
\caption[Passive fraction of the C\--GaND]{Passive, or quiescent, fraction of the number
density selected samples selected at three different number densities: $n=10^{-4}\rm{Mpc^{-3}}$ (red boxes), $n=10^{-5}\rm{Mpc^{-3}}$ (blue triangles) and $n=3\times10^{-4}\rm{Mpc^{-3}}$ (black solid circles) vs redshift. The points at each number density selection denote the fraction of galaxies selected as passive via the $UVJ$ method. The solid lines show the best fit to the passive fractions with the form of equation 9 at the three number densities with the colour of this line
corresponding to the number density.  The errors on our fractions are derived from Monte Carlo analysis. The dotted line at the bottom with
open circle points is the passive fraction found by Papovich
et al. (2015) for M31 mass galaxies, and the dashed line with the
crosses for points shows the passive evolution for galaxies
selected with abundance matching at masses log M$_{*} > 10^{11.8}$ by
Marchesini et al. (2014). }
\vspace{0cm}
\label{pass} 
 \end{figure}

\begin{equation}
F_{passive}= 1 - (0.05\pm0.02) \times \rm{e}^{(1.0\pm0.2) \times z}.
\end{equation}

\noindent We find that the passive fraction of progenitor galaxies for
this selection undergoes a 
significant evolution over the redshift range $0.3<z<3.0$. Within our 
lowest redshift bin at $z \sim 0.5$, $94\pm8\%$ of the progenitor galaxies 
are passive, much like their local universe counterparts. In our highest 
redshift bin $\sim 50\%$ of the progenitor galaxies are passive by $z \sim 2.5$. This 
implies that about half of the progenitors of today's massive galaxies had 
already stopped actively star forming by $z=3.0$. 

We find a similar trend for the other number densities used in this study, 
$n = 10^{-5}$~Mpc$^{-3}$ and $n = 3 \times 10^{-4}$~Mpc$^{-3}$,
where we also find that the passive fraction is near 90\% by $z = 1.5$.  
In fact there does not appear to be a strong dependence on stellar mass, or number
density selection, in how the fraction of galaxies which are passive evolve
with time. This result however could easily reside in the uncertainties
which arise from determining passive fractions. However, 
there is a trend such that on average the higher mass 
and lower number density selected objects have a higher passive fraction
at all redshifts.  

This indicates that the average progenitor of local massive galaxies have 
had a red rest-frame colour since $z=3$, although with a large
scatter and with some blue galaxies. 
  This is similar to, but slightly different, to the
findings of \cite{Marchesini2014}, 
who find that the progenitors of the local ultra-massive galaxies 
(with $\rm{log(}M_*/M_\odot)=11.8$) have blue average rest frame 
colours and $\sim17\%$ are selected as passive at $z>2.5$ (Figure~9). 
Papovich et al. (2015) find an even lower fraction which is passive. Our
result is higher in passive fraction likely due to the fact that
our galaxies are more massive and thus more likely to be passive
up to $z \sim 3$.   However, this does not explain the tension with 
Marchesini et al. (2014) and there thus remains an unexplained
inconsistency between our results and theirs.

The observed weakening of the colour-density relation at $z>2$ 
(e.g. \citealt{Chuter2011}, \citealt{Grutzbauch2011}) implies that the 
environments of galaxies have not been fully established at high redshift. 
Therefore, the role of environmental quenching mechanisms, such as ram 
pressure stripping, are unlikely to play a dominant role in the quenching of 
the progenitors of local massive galaxies at early cosmic times. The 
result that we present here shows that a large fraction of galaxies are 
already passive by $z=3.0$ and implies that internal quenching mechanisms, 
such as the hot halo model, are likely responsible.

\subsection{S\'{e}rsic Index and Size Evolution}

\begin{figure} 
\subfloat[]{\includegraphics[scale=0.5]{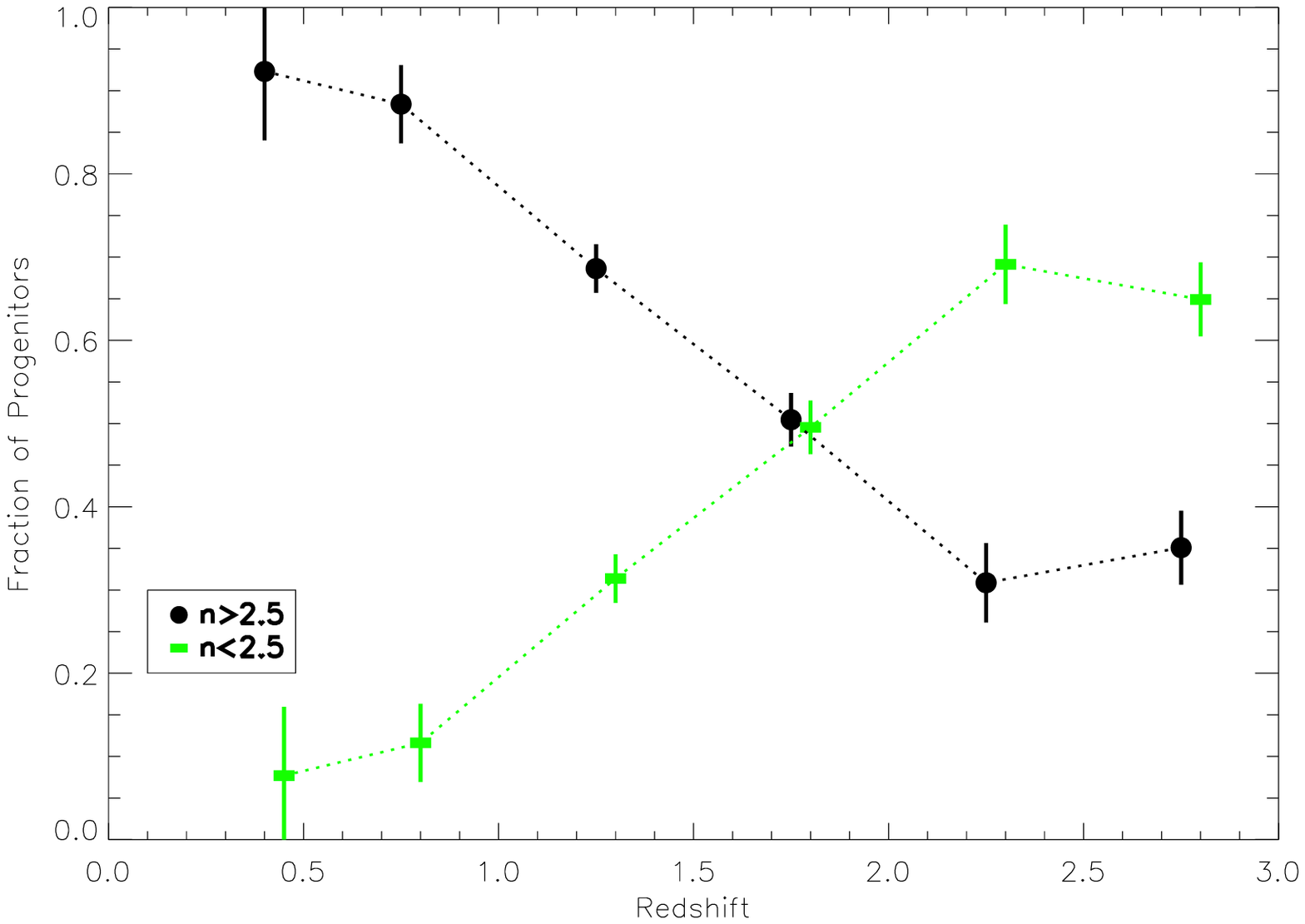}}\\
\subfloat[]{\includegraphics[scale=0.5]{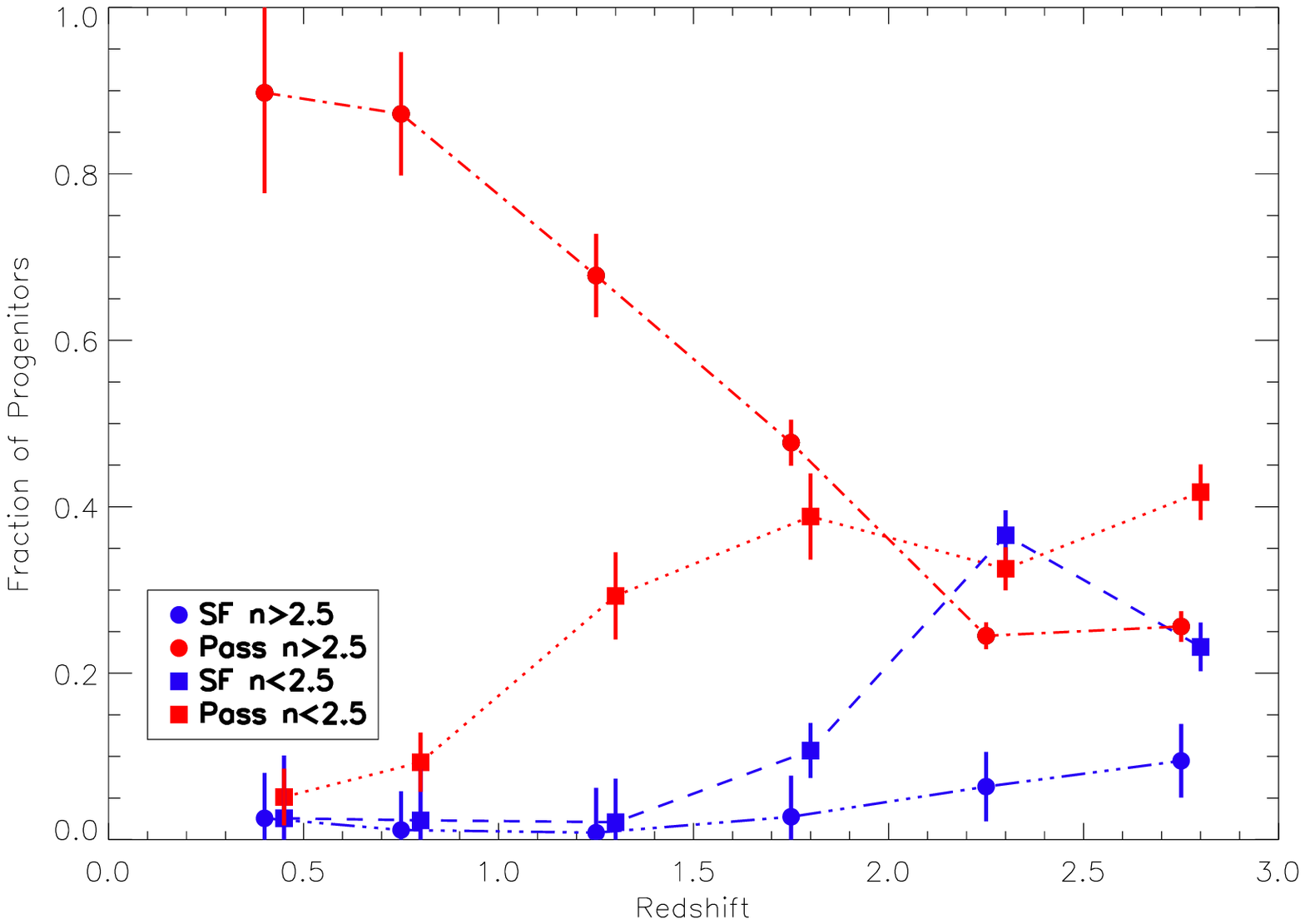}}
\caption[The fractions of the progenitors of local massive galaxies with high ($n>2.5$) and low ($n<2.5$) S\'{e}rsic indices.]{Fraction of the progenitors of local massive galaxies with high ($n>2.5$) and low ($n<2.5$) S\'{e}rsic indices. Figure (a) show the evolution of the whole progenitor sample with, green rectangles showing the fraction of progenitors with low S\'{e}rsic index and, black circles showing progenitors with high S\'{e}rsic light profiles as a function of redshift. Error bars are derived using Monte Carlo analysis. Figure (b) shows the high and low S\'{e}rsic populations split into star forming and passive systems.}
 \label{n}
 \end{figure} 

The present day massive galaxy population is dominated by objects with 
early-type morphologies and high S\'{e}rsic indices (e.g. 
\citealt{Baldry2004}, \citealt{Conselice2006}, \citealt{Buitrago2013}). 
Examining similar stellar mass objects at $z>2$ studies have found this 
not to be the case (e.g., \citealt{Mortlock2013}, \citealt{Buitrago2013}, 
\citealt{Bruce2014}). However, this has not been examined using a 
number density selected sample.   Avoiding the precursor bias is critical,
as when we compare a stellar mass selected sample we are obtaining samples
which can be 95\% or more different between
$z = 3$ and $z = 0$ (e.g., Mundy et al. 2015).

In Figure~10a we show the evolution of the S\'{e}rsic indices of 
the progenitors of local massive galaxies at our measured
density of $n = 10^{-4}$ Mpc$^{-3}$. The progenitor galaxies have 
been split into high and low S\'{e}rsic index systems with a dividing line 
at $n=2.5$. The value of $n=2.5$ has been used in many studies as a 
quantitative way to segregate between early and late type galaxies, with 
early type galaxies having $n>2.5$, although for mass selected 
samples (e.g. \citealt{Shen2003}, \citealt{Barden2005}, 
\citealt{Mcintosh2005}, \citealt{Buitrago2013}). The fraction of 
progenitors with high S\'{e}rsic indices is represented by the green 
rectangles, and the fraction with low S\'{e}rsic indices is represented 
by the black circles. Figure~10a clearly indicates that the fraction of 
the progenitors of local massive galaxies with lower S\'{e}rsic indices 
has greatly increased with redshift, with $8\pm5\%$ of the progenitor 
galaxies at $z=0.3$ having low  S\'{e}rsic indices, increasing to 
$65\pm7\%$ at $z=3$.  We find essentially the same results at the
other two number densities considered in the Appendix.

If we take the assumption that objects with low S\'{e}rsic indices have 
a disk-like morphology, this result implies that the progenitor galaxies at 
high redshift are mostly disky galaxies. However, this assumption breaks 
down if we consider the effect of galaxies with disturbed and irregular 
morphologies. Both \cite{Buitrago2013}  and \cite{Mortlock2013} showed 
that galaxies at high redshift with low S\'{e}rsic indices also display 
disturbed morphologies when examined using visual classification. These 
studies also showed that the number of galaxies with disturbed visual 
morphologies increases dramatically with redshift, with $\sim40\%$ of 
massive galaxies showing a disturbed  morphology at $z=3$. This increase in 
the number of galaxies with disturbed morphologies could be linked to the 
increase in the importance of major mergers with redshift shown in e.g.,
Bluck et al. (2012); \cite{Ownsworth2014}, Conselice et al. (2003, 2009, 2011).

Therefore, using just the S\'{e}rsic profile information we cannot 
determine if the progenitor galaxy population we present here are true 
disks or disturbed galaxies. Examining the asymmetries of
our galaxies (e.g. CAS \citealt{Conselice2006CAS}) shows that a large
fraction of these systems are indeed undergoing mergers (see 
\citealt{Mortlock2013} for more discussion).  From the 
limited number of galaxies at these masses studied in \cite{Mortlock2013} 
we know that at $z > 2.75$ around 60\% of these galaxies are peculiar with the remainder having a visual
morphology of smooth elliptical like systems.  The fact that these galaxies
visually do not have disks shows that they are not the equivalent of lower
redshift disks and therefore have a different formation history.

Also from Figure~10a we find that the redshift where the 
progenitor galaxies transition into the high S\'{e}rsic index dominated 
population we see in the local Universe is between $1.5<z<2.0$. This is in 
agreement with previous studies which examine the morphological change 
of galaxies with similar stellar masses as our sample 
(e.g. \citealt{Mortlock2013}).  \cite{Mortlock2013} show that this is also the 
redshift range where peculiar
galaxies transition into normal ellipticals and spirals.  We see the same
for our galaxies but within a transition from low to high S\'{e}rsic index
systems.

We further divide the high and low S\'{e}rsic progenitor samples into star 
forming and passive systems using our $UVJ$ selection and present the results 
in Figure~10b. This figure shows the clear dominance at $z<1.7$ of 
the passive high S\'{e}rsic index systems that we associate with massive 
galaxies in the local Universe. The population of high S\'{e}rsic index 
galaxies is, at all redshifts examined in this study, dominated by the 
passive population. The star forming and high S\'{e}rsic index systems are the
least abundant at $z=3.0$, only constituting $9\pm3\%$ of the total 
progenitor population at this redshift. 

Examining the low S\'{e}rsic index systems we find 
that these objects are dominated at almost all redshifts by passive systems, 
much like the high S\'{e}rsic index population. At $z=2.5$,  $41\pm4\%$ of 
the progenitor galaxies are passive and have low S\'{e}rsic indices, and 
$23\pm3\%$ of the progenitor galaxies are star forming and have low S\'{e}rsic 
indices. This result implies that passive low S\'{e}rsic index systems out 
number star forming low S\'{e}rsic index systems by nearly a factor of 
two within the progenitor massive galaxy population. This result is 
surprising as the morphologies that constitute the low S\'{e}rsic index 
population are generally thought to be star forming. However, this result 
is in agreement with recent work by \cite{Bruce2014} using two component 
light profile fitting which has shown that a large fraction, 
$\sim38\%$ of passive massive galaxies at $z>1.5$ are disk-like dominant 
systems.

We also find in our previous work on this sample (Ownsworth et al. 2014) that
the size evolution for this population evolves by a factor of at most two
between $0.5 < z < 3$.  This is much less than the factor of five or so found
for a mass selected sample over the same redshift range (e.g., Buitrago
et al. 2008).  However, there is still an evolution in size that must come about
from something other than finding the wrong progenitors of low redshift
massive galaxies.

A clue to this is that the dominant population at high-z for this sample
are galaxies with low S\'{e}rsic indices, which become larger at higher
redshifts. Also, as we discuss above, the dominant type of galaxy in this
selection at $z \sim 3$ are passive systems.  Therefore this shows that star
formation is unlikely to be the cause of the increased sizes and higher
S\'{e}rsic indices as time goes on (see also Ownsworth et al. 2012).  The
most likely explanation for this is minor mergers which occur in about the
right quantity to account for the evolution seen (e.g., Bluck et al. 2012).
The fact that there is a decoupling between when galaxies become passive 
(which occurs at $z > 1$)
and this structural evolution implies that a dynamical effect, such as these
minor mergers, are the cause of the continued structural evolution at $z < 1$.

Furthermore, we see a strong correlation between the galaxies which have a high
S\'{e}rsic index and those that are passive.  This demonstrates that passivity
and structure are highly correlated, and that most likely the formation of
a galaxy into a concentrated profile leads to the shutting down of star 
formation.

\section{Summary}

In this paper we present a study of the evolution of the properties of 
galaxies constituting a constant number density selected sample over the 
redshift range of $0.3<z<3.0$.  Our main sample are galaxies selected
by a constant number density of $n= 10^{-4}$ Mpc$^{-3}$ which ranges from
galaxies with stellar masses M$_{*} = 10^{10.54}$ $\rm{M_{\odot}}$
at $z = 2.75$ to M$_{*} = 10^{11.24}$ $\rm{M_{\odot}}$ at $z = 0.3$. 
We examine the evolution of  colour, location on the colour-stellar mass 
diagram, passivity and, structural 
parameters. We find the following principle results:

\begin{itemize}
\item We find that the average $U-V$ and $V-J$ colours of the progenitors 
of local massive galaxies have been located within the $UVJ$ defined passive 
region since at least $z=3.0$. However the progenitors that are classified as 
star forming have a large scatter in both colours, and in some cases show 
redder colours than the passive galaxies at the same epoch. When we examine 
these galaxies using the colour-stellar mass diagram we also find that the 
average progenitor of local massive galaxies has not lived on the blue cloud 
since $z=3.0$. Using stellar population models we find that the
progenitor galaxies which are passive 
have old stellar ages (age $> 5$ Gyr) and appear to 
show hints that they have been passively evolving since at least $z=4$.

\item We examine how the progenitor population becomes the passive 
population we see today over this redshift range. We find that the passive 
fraction of the progenitor galaxies undergoes significant evolution from 
$z=3.0$, increasing from 27\% at $z=3.0$ (from our fit, 50\% observed at 
$z=2.5$) 
to $94\pm8\%$ at $z=0.3$. 
This implies that over half of the population of the progenitors of local 
massive galaxies have already stopped forming stars by $z=3.0$. Also the 
star formation density of the progenitors shows signs of galaxy  
downsizing, with galaxies selected in a given number density becoming
passive earlier than those selected at higher densities.

\item The morphological evolution of the progenitor galaxies is 
probed using the evolution of the S\'{e}rsic indices within the sample. 
We find that these galaxies are dominated at high redshifts by low S\'{e}rsic 
index ($n<2.5$) light profiles and evolve to be come high S\'{e}rsic index 
($n>2.5$) dominated objects by $z=1.7$. We further split the high and 
low S\'{e}rsic populations into star forming and passive systems. We 
find that passive high S\'{e}rsic index systems are the most abundant 
objects at $z<1.7$, equivalent to their descendants at $z \sim 0$. There 
exists a small population of star forming high S\'{e}rsic index objects at 
high redshift but they rapidly decrease towards low redshift. We also find 
that $41\pm4\%$ of the population within the highest redshift bin are 
passive low S\'{e}rsic index objects. This could imply that a significant 
proportion of the progenitors of massive galaxies were passive disk-like systems at 
early times. However, this low S\'{e}rsic index trend could be  
driven by the increase in the abundance of morphologically disturbed
 systems at higher redshifts.

We previously investigated the size evolution of the constant number density 
selected sample using no passivity cuts and find that the sizes of the 
progenitors of massive galaxies range from a factor of $1.8$ to $1.2$ 
smaller than local early type galaxies of similar mass over $0.5 < z < 3$
(Ownsworth et al. 2014). This is smaller than previous studies have found, quoting size evolution factors of two to four.   

However, this does show that galaxy evolution is occurring within the galaxy 
population selected to be as close as possible the same systems over redshifts.
Previous studies were limited in making this conclusion based on using a
constant mass selection whereby there is significant precursor bias
contamination to the level of 95\% since $z < 1$.  

\end{itemize}

To further this work, especially when probing high number densities, or lower
mass objects, or to extend to $z > 3$ will require  larger and deeper surveys. Future 
telescopes such as JWST, E-ELT and Euclid will be able to push these 
trends out to higher redshifts and be able to investigate the full history 
of local massive galaxies.  However, in all surveys any future conclusions about galaxy
evolution must use a number density selected sample as outlined in this paper,
or else will have a significant and catastrophic precursor bias.

\section*{Acknowledgments}
We thank the anonymous referee for critical and useful comments which improved this paper significantly. We also thank the UDS team for their support and work on this survey and assistance on this paper. We acknowledge funding from the STFC and the Leverhulme trust for supporting this work. A.M. acknowledges funding from the STFC and a European Research Council Consolidator Grant (PI R. McLure).

\bibliographystyle{mnras}
\bibliography{refs}{}

\begin{thebibliography}{54}
\expandafter\ifx\csname natexlab\endcsname\relax\def\natexlab#1{#1}\fi

\bibitem[{Baldry} et~al.(2006){Baldry}, {Balogh}, {Bower} et~al.]{Baldry2006}
{Baldry} I.~K., {Balogh} M.~L., {Bower} R.~G., et~al., 2006, \mnras, 373, 469

\bibitem[{Baldry} et~al.(2004){Baldry}, {Glazebrook}, {Brinkmann}
  et~al.]{Baldry2004}
{Baldry} I.~K., {Glazebrook} K., {Brinkmann} J., et~al., 2004, \apj, 600, 681

\bibitem[{Barden} et~al.(2012){Barden}, {H{\"a}u{\ss}ler}, {Peng}, {McIntosh}
  \& {Guo}]{Barden2012}
{Barden} M., {H{\"a}u{\ss}ler} B., {Peng} C.~Y., {McIntosh} D.~H., {Guo} Y.,
  2012, \mnras, 422, 449

\bibitem[{Barden} et~al.(2005){Barden}, {Rix}, {Somerville} et~al.]{Barden2005}
{Barden} M., {Rix} H.-W., {Somerville} R.~S., et~al., 2005, \apj, 635, 959

\bibitem[{Bell} et~al.(2003){Bell}, {McIntosh}, {Katz} \& {Weinberg}]{Bell2003}
{Bell} E.~F., {McIntosh} D.~H., {Katz} N., {Weinberg} M.~D., 2003, \apjs, 149,
  289

\bibitem[{Bower} et~al.(1992){Bower}, {Lucey} \& {Ellis}]{Bower1992}
{Bower} R.~G., {Lucey} J.~R., {Ellis} R.~S., 1992, \mnras, 254, 601

\bibitem[{Brammer} et~al.(2008){Brammer}, {van Dokkum} \& {Coppi}]{EAZY2008}
{Brammer} G.~B., {van Dokkum} P.~G., {Coppi} P., 2008, \apj, 686, 1503

\bibitem[{Bruce} et~al.(2014){Bruce}, {Dunlop}, {McLure} et~al.]{Bruce2014}
{Bruce} V.~A., {Dunlop} J.~S., {McLure} R.~J., et~al., 2014, ArXiv e-prints

\bibitem[{Bruzual} \& {Charlot}(2003)]{Bruzual2003}
{Bruzual} G., {Charlot} S., 2003, \mnras, 344, 1000

\bibitem[{Buitrago} et~al.(2013){Buitrago}, {Trujillo}, {Conselice} \&
  {H{\"a}u{\ss}ler}]{Buitrago2013}
{Buitrago} F., {Trujillo} I., {Conselice} C.~J., {H{\"a}u{\ss}ler} B., 2013,
  \mnras, 428, 1460

\bibitem[{Charlot} \& {Fall}(2000)]{Charlot2000}
{Charlot} S., {Fall} S.~M., 2000, \apj, 539, 718

\bibitem[{Chuter} et~al.(2011){Chuter}, {Almaini}, {Hartley}
  et~al.]{Chuter2011}
{Chuter} R.~W., {Almaini} O., {Hartley} W.~G., et~al., 2011, \mnras, 413, 1678

\bibitem[{Cole} et~al.(2001){Cole}, {Norberg}, {Baugh} et~al.]{Cole2001}
{Cole} S., {Norberg} P., {Baugh} C.~M., et~al., 2001, \mnras, 326, 255

\bibitem[{Conselice}(2006{\natexlab{a}})]{Conselice2006}
{Conselice} C.~J., 2006{\natexlab{a}}, \apj, 638, 686

\bibitem[{Conselice}(2006{\natexlab{b}})]{Conselice2006CAS}
{Conselice} C.~J., 2006{\natexlab{b}}, \mnras, 373, 1389

\bibitem[{Conselice} et~al.(2011){Conselice} et al. ]{C2011}
{Conselice} C.~J., et al. 2011, \mnras, 417, 2770

\bibitem[{Conselice} et~al.(2013){Conselice}, {Mortlock}, {Bluck},
  {Gr{\"u}tzbauch} \& {Duncan}]{C2013}
{Conselice} C.~J., {Mortlock} A., {Bluck} A.~F.~L., {Gr{\"u}tzbauch} R.,
  {Duncan} K., 2013, \mnras, 430, 1051

\bibitem[{Daddi} et~al.(2004){Daddi}, {Cimatti}, {Renzini} et~al.]{Daddi2004}
{Daddi} E., {Cimatti} A., {Renzini} A., et~al., 2004, \apj, 617, 746

\bibitem[{Daddi} et~al.(2007){Daddi}, {Dickinson}, {Morrison}
  et~al.]{Daddi2007}
{Daddi} E., {Dickinson} M., {Morrison} G., et~al., 2007, \apj, 670, 156

\bibitem[{Duncan} et~al.(2014){Duncan}, {Conselice}, {Mortlock}
  et~al.]{Duncan2014}
{Duncan} K., {Conselice} C.~J., {Mortlock} A., et~al., 2014, \mnras, 444, 2960

\bibitem[{Fischera} \& {Dopita}(2005)]{Fischera2005}
{Fischera} J., {Dopita} M., 2005, \apj, 619, 340

\bibitem[{Furusawa} et~al.(2008){Furusawa}, {Kosugi}, {Akiyama}, {Takata},
  {Sekiguchi} \& {Furusawa}]{Furusawa2008}
{Furusawa} H., {Kosugi} G., {Akiyama} M., {Takata} T., {Sekiguchi} K.,
  {Furusawa} J., 2008, 399, 131

\bibitem[{Gallazzi} et~al.(2005){Gallazzi}, {Charlot}, {Brinchmann}, {White} \&
  {Tremonti}]{Gallazzi2005}
{Gallazzi} A., {Charlot} S., {Brinchmann} J., {White} S.~D.~M., {Tremonti}
  C.~A., 2005, \mnras, 362, 41

\bibitem[{Grogin} et~al.(2011){Grogin}, {Kocevski}, {Faber} et~al.]{Grogin2011}
{Grogin} N.~A., {Kocevski} D.~D., {Faber} S.~M., et~al., 2011, \apjs, 197, 35

\bibitem[{Gr{\"u}tzbauch} et~al.(2011){Gr{\"u}tzbauch}, {Conselice}, {Bauer}
  et~al.]{Grutzbauch2011}
{Gr{\"u}tzbauch} R., {Conselice} C.~J., {Bauer} A.~E., et~al., 2011, \mnras,
  418, 938

\bibitem[{Hartley} et~al.(2013){Hartley}, {Almaini}, {Mortlock}
  et~al.]{Hartley2013}
{Hartley} W.~G., {Almaini} O., {Mortlock} A., et~al., 2013, \mnras, 431, 3045

\bibitem[{Hopkins} \& {Beacom}(2006)]{Hopkins2006}
{Hopkins} A.~M., {Beacom} J.~F., 2006, \apj, 651, 142

\bibitem[{Kauffmann} et~al.(2003){Kauffmann}, {Heckman}, {White}
  et~al.]{Kauffmann2003}
{Kauffmann} G., {Heckman} T.~M., {White} S.~D.~M., et~al., 2003, \mnras, 341,
  54

\bibitem[{Kaviraj} et~al.(2013){Kaviraj}, {Cohen}, {Ellis} et~al.]{Kaviraj2013}
{Kaviraj} S., {Cohen} S., {Ellis} R.~S., et~al., 2013, \mnras, 428, 925

\bibitem[{Kennicutt}(1983)]{Kennicutt1983}
{Kennicutt} Jr. R.~C., 1983, \apj, 272, 54

\bibitem[{Kennicutt}(1998)]{Kennicutt1998}
{Kennicutt} Jr. R.~C., 1998, \araa, 36, 189

\bibitem[{Koekemoer} et~al.(2011){Koekemoer}, {Faber}, {Ferguson}
  et~al.]{Koekemoer2011}
{Koekemoer} A.~M., {Faber} S.~M., {Ferguson} H.~C., et~al., 2011, \apjs, 197,
  36

\bibitem[{Lani} et~al.(2013){Lani}, {Almaini}, {Hartley} et~al.]{Lani2013}
{Lani} C., {Almaini} O., {Hartley} W.~G., et~al., 2013, \mnras, 435, 207

\bibitem[{Marchesini} et~al.(2014){Marchesini}, {Muzzin}, {Stefanon}
  et~al.]{Marchesini2014}
{Marchesini} D., {Muzzin} A., {Stefanon} M., et~al., 2014, ArXiv e-prints

\bibitem[{McIntosh} et~al.(2005){McIntosh}, {Bell}, {Rix} et~al.]{Mcintosh2005}
{McIntosh} D.~H., {Bell} E.~F., {Rix} H.-W., et~al., 2005, \apj, 632, 191

\bibitem[{McLure} et~al.(2013){McLure}, {Dunlop}, {Bowler} et~al.]{McLure2013}
{McLure} R.~J., {Dunlop} J.~S., {Bowler} R.~A.~A., et~al., 2013, \mnras, 432,
  2696

\bibitem[{Meurer} et~al.(1999){Meurer}, {Heckman} \& {Calzetti}]{Meurer1999}
{Meurer} G.~R., {Heckman} T.~M., {Calzetti} D., 1999, \apj, 521, 64

\bibitem[{Mortlock} et~al.(2013){Mortlock}, {Conselice}, {Hartley}
  et~al.]{Mortlock2013}
{Mortlock} A., {Conselice} C.~J., {Hartley} W.~G., et~al., 2013, \mnras, 433,
  1185

\bibitem[{Muzzin} et~al.(2013){Muzzin}, {Marchesini}, {Stefanon}
  et~al.]{Muzzin2013}
{Muzzin} A., {Marchesini} D., {Stefanon} M., et~al., 2013, \apj, 777, 18

\bibitem[{Ownsworth} et~al.(2014){Ownsworth}, {Conselice}, {Mortlock}
  et~al.]{Ownsworth2014}
{Ownsworth} J.~R., {Conselice} C.~J., {Mortlock} A., et~al., 2014, \mnras, 445,
  2198

\bibitem[{Ownsworth} et~al.(2012){Ownsworth}, {Conselice}, {Mortlock},
  {Hartley} \& {Buitrago}]{O2012}
{Ownsworth} J.~R., {Conselice} C.~J., {Mortlock} A., {Hartley} W.~G.,
  {Buitrago} F., 2012, \mnras, 426, 764

\bibitem[{Papovich} et~al.(2011){Papovich}, {Finkelstein}, {Ferguson}, {Lotz}
  \& {Giavalisco}]{Papovich2011}
{Papovich} C., {Finkelstein} S.~L., {Ferguson} H.~C., {Lotz} J.~M.,
  {Giavalisco} M., 2011, \mnras, 412, 1123

\bibitem[{Papovich} et~al.(2015){Papovich} et al. ]{Papovich2015}
{Papovich} C. et al. 2015, \mnras, 803, 26


\bibitem[{Patel} et~al.(2013){Patel}, {van Dokkum}, {Franx} et~al.]{Patel2013}
{Patel} S.~G., {van Dokkum} P.~G., {Franx} M., et~al., 2013, \apj, 766, 15

\bibitem[{Peng}(2010)]{Peng2010}
{Peng} C., 2010, in { American Astronomical Society Meeting Abstracts \#215\/},
  vol.~42 of { Bulletin of the American Astronomical Society\/},  229.09

\bibitem[{Pozzetti} et~al.(2010){Pozzetti}, {Bolzonella}, {Zucca}
  et~al.]{Pozzetti2010}
{Pozzetti} L., {Bolzonella} M., {Zucca} E., et~al., 2010, \aap, 523, A13

\bibitem[{Prevot} et~al.(1984){Prevot}, {Lequeux}, {Prevot}, {Maurice} \&
  {Rocca-Volmerange}]{Prevot1984}
{Prevot} M.~L., {Lequeux} J., {Prevot} L., {Maurice} E., {Rocca-Volmerange} B.,
  1984, \aap, 132, 389

\bibitem[{Renzini}(2006)]{Renzini2006}
{Renzini} A., 2006, \araa, 44, 141

\bibitem[{S\'{e}rsic}(1968)]{sersic1968}
{S\'{e}rsic} J.~L., 1968, {Atlas de galaxias australes}

\bibitem[{Shen} et~al.(2003){Shen}, {Mo}, {White} et~al.]{Shen2003}
{Shen} S., {Mo} H.~J., {White} S.~D.~M., et~al., 2003, \mnras, 343, 978

\bibitem[{Simpson} et~al.(2006){Simpson}, {Mart{\'{\i}}nez-Sansigre},
  {Rawlings} et~al.]{Simpson2006}
{Simpson} C., {Mart{\'{\i}}nez-Sansigre} A., {Rawlings} S., et~al., 2006,
  \mnras, 372, 741

\bibitem[{Taylor-Mager} et~al.(2007){Taylor07},et~al.]{Taylor2007}
{Taylor-Mager} V.A., {Conselice} C.J., {Windhorst} R.A., {Jansen} R.A. 2007, ApJ, 569, 162


\bibitem[{Taylor} et~al.(2014){Taylor}, {Hopkins}, {Baldry} et~al.]{Taylor2014}
{Taylor} E.~N., {Hopkins} A.~M., {Baldry} I.~K., et~al., 2014, ArXiv e-prints

\bibitem[{Ueda} et~al.(2008){Ueda}, {Watson}, {Stewart} et~al.]{Ueda2008}
{Ueda} Y., {Watson} M.~G., {Stewart} I.~M., et~al., 2008, \apjs, 179, 124

\bibitem[{van der Wel} et~al.(2012){van der Wel}, {Bell}, {H{\"a}ussler}
  et~al.]{Wel2012}
{van der Wel} A., {Bell} E.~F., {H{\"a}ussler} B., et~al., 2012, \apjs, 203, 24

\bibitem[{Whitaker} et~al.(2012){Whitaker}, {Kriek}, {van Dokkum}
  et~al.]{Whitaker2012b}
{Whitaker} K.~E., {Kriek} M., {van Dokkum} P.~G., et~al., 2012, \apj, 745, 179

\bibitem[{Williams} et~al.(2009){Williams}, {Quadri}, {Franx}, {van Dokkum} \&
  {Labb{\'e}}]{Williams2009}
{Williams} R.~J., {Quadri} R.~F., {Franx} M., {van Dokkum} P., {Labb{\'e}} I.,
  2009, \apj, 691, 1879

\end{thebibliography}
\appendix
\section{Appendix}

In the Appendix we show the $UVJ$ colours of a few other number density 
selections to compare with the nominal one at $n=10^{-4}\, \rm{Mpc^{-3}}$. We
investigate this to determine how our results depend upon the selection
method using other number
densities.

\begin{figure*}
\includegraphics[scale=0.80]{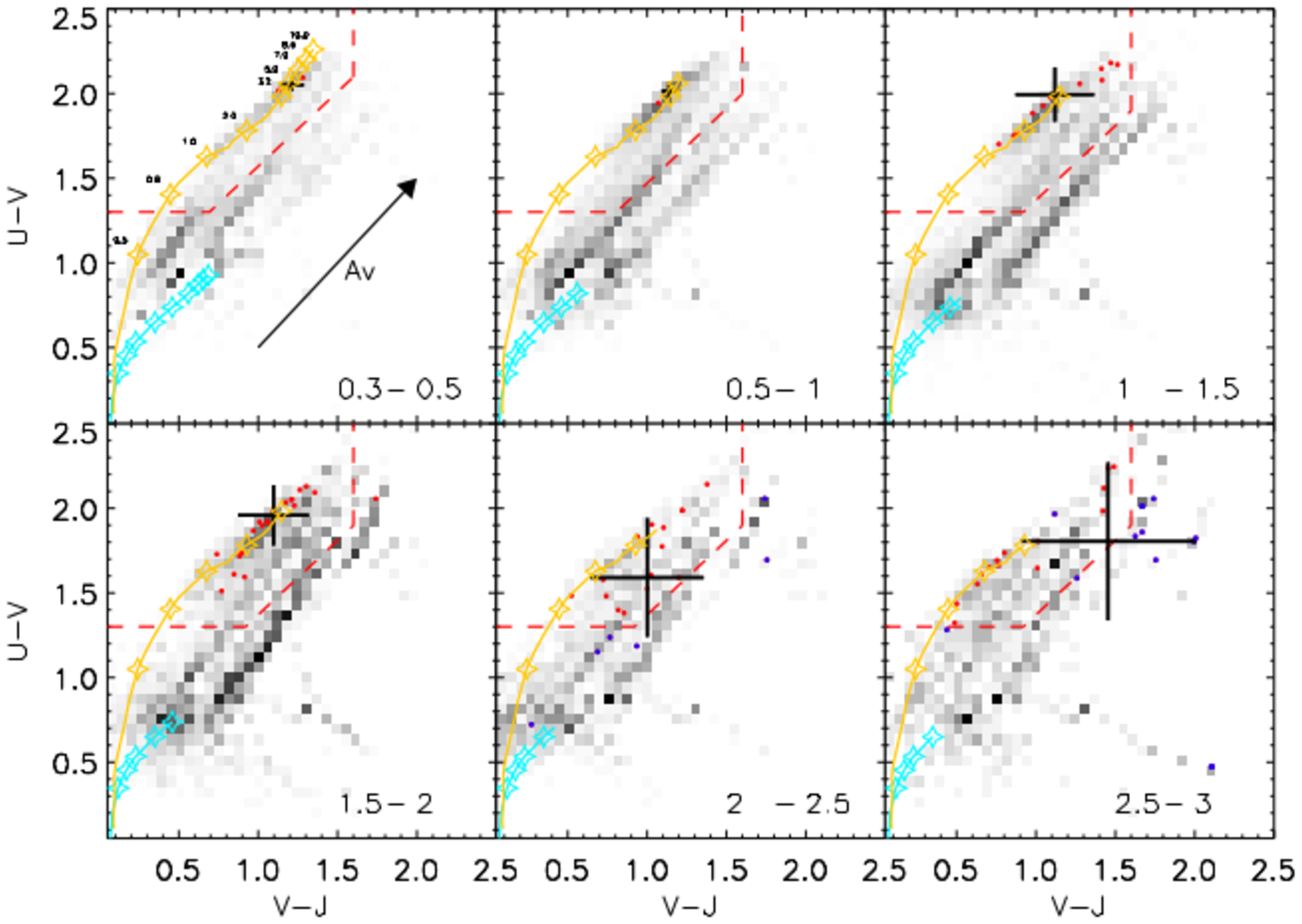}      
\caption[Rest frame $U-V$ versus $V-J$ diagram in redshift bins between $z=0.3$ and $z=3.0$ of the constant number density selected sample]{Rest frame $U-V$ versus $V-J$ diagram in redshift bins between $z=0.3$ and $z=3.0$ for the constant number density selected sample with $n=10^{-5}\rm{Mpc^{-3}}$, corresponding to a mass limit of log M$_{*} \sim  11.59$ at $z \sim 0.3$. The red dashed line denotes the $UVJ$ passive selection. Red circles show the progenitors of massive galaxies that are selected as passive via the $UVJ$ method. Blue circles show the progenitors of massive galaxies that are selected as star forming via the $UVJ$ method and $24\mu m $ criteria. The black cross shows the median colour and standard deviation for the progenitor sample in each redshift bin. The colour evolution tracks from \protect \cite{Bruzual2003} SSP models are also shown. The light blue line shows a constant star formation history with no dust and the yellow line shows an exponentially declining star formation history with $\tau=0.1\,\rm{Gyr}$ (see Figure~2). The open stars represent model colours at the specified ages, given in Gyr. The colour evolution tracks are plotted up to the  age of the Universe in each redshift bin.}
\label{UVJhigh}   
\end{figure*}

\begin{figure*}
\includegraphics[scale=0.65]{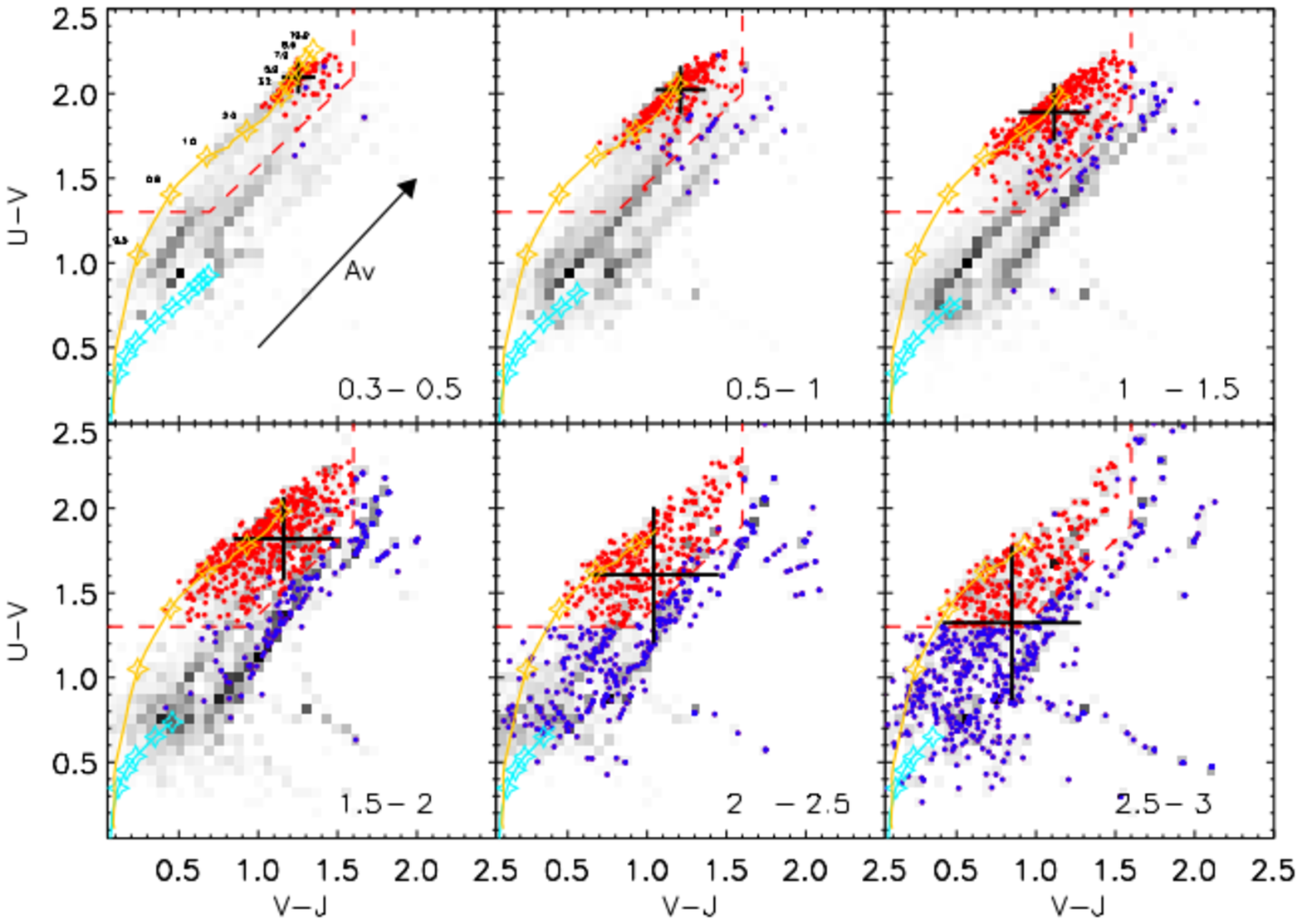}  
\caption[Rest frame $U-V$ versus $V-J$ diagram in redshift bins between $z=0.3$ and $z=3.0$ of the constant number density selected sample]{Rest frame $U-V$ versus $V-J$ diagram in redshift bins between $z=0.3$ and $z=3.0$ for the constant number density selected sample with $n=3\times10^{-4}\rm{Mpc^{-3}}$, corresponding to a mass limit of log M$_{*} \sim  10.93$ at $z \sim 0.3$. The red dashed line denotes the $UVJ$ passive selection. Red circles show the progenitors of massive galaxies that are selected as passive via the $UVJ$ method. Blue circles show the progenitors of massive galaxies that are selected as star forming via the $UVJ$ method and $24\mu m $ criteria. The black cross shows the median colour and standard deviation for the progenitor sample in each redshift bin. The colour evolution tracks from \protect \cite{Bruzual2003} SSP models are also shown. The light blue line shows a constant star formation history with no dust and the yellow line shows an exponentially declining star formation history with $\tau=0.1\,\rm{Gyr}$ (see Figure~2). The open stars represent model colours at the specified ages, given in Gyr. The colour evolution tracks are plotted up to the  age of the Universe in each redshift bin.}
\label{UVJlow} 
\end{figure*}

The other number densities that we investigate are: 
 $n=10^{-5}\, \rm{Mpc^{-3}}$ 
(Figure \ref{UVJhigh}, Figure~A3) and $n=3\times10^{-4}\, \rm{Mpc^{-3}}$ 
(Figure \ref{UVJlow}, Figure~A4). Both of these number density selections show 
similar behavior as the $n=10^{-4}\, \rm{Mpc^{-3}}$ sample (\S 3.1).   The only
difference is that we see a signature of downsizing, whereby
the lower number densities, tracing higher mass galaxies, become a homogeneous
red/passive population earlier than galaxies selected at higher number
densities, or lower mass galaxies (see main body of paper for details.)

\begin{figure*} 
\includegraphics[scale=0.65]{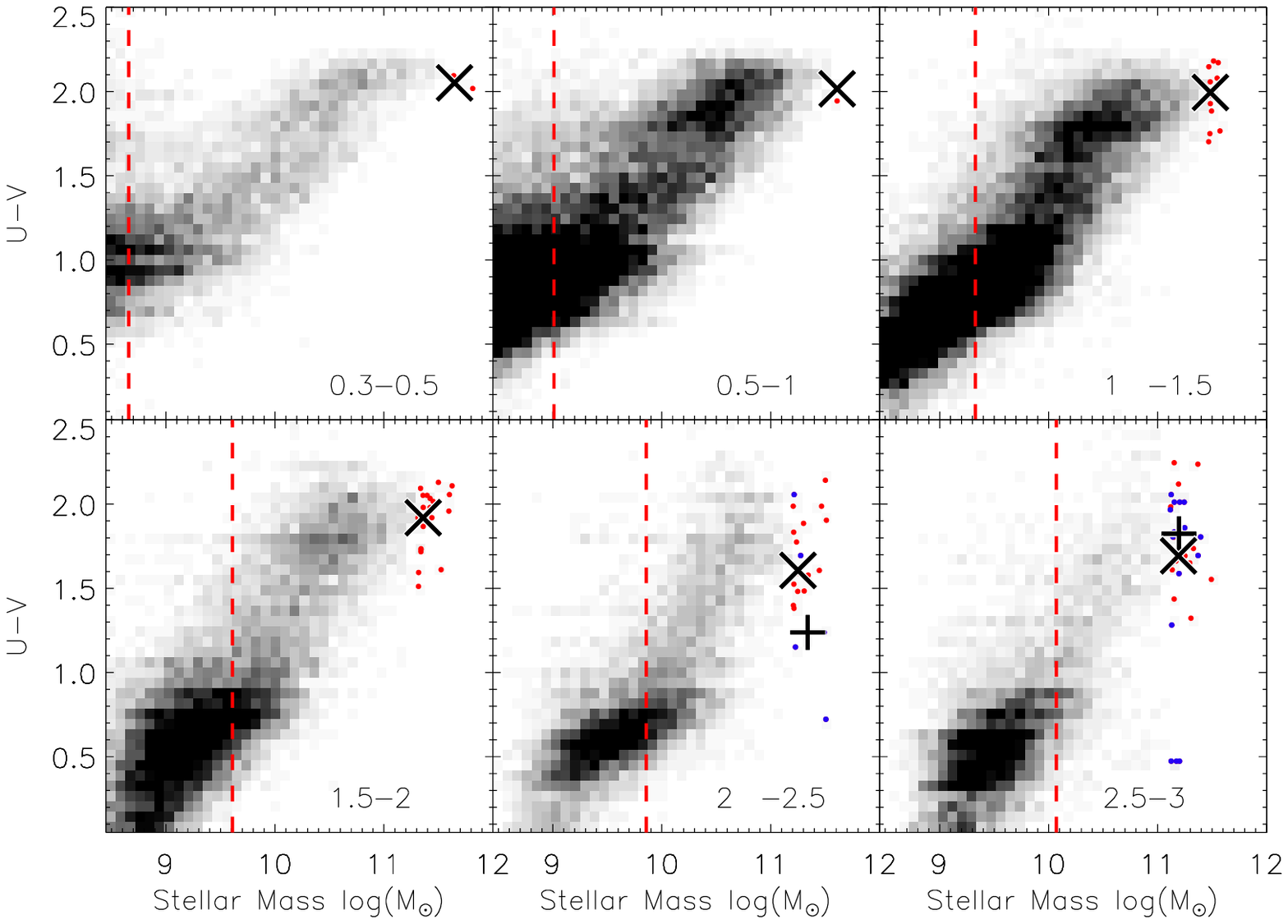}     
\caption[Stellar mass versus rest frame $U-V$ colour for all galaxies selected via the ]{Stellar mass versus rest frame $U-V$ colour for all galaxies selected via the constant number density selected sample with $n=10^{-5}\rm{Mpc^{-3}}$. The red circles show the progenitors of massive galaxies that are selected as passive via the $UVJ$ method. The blue circles show the progenitors of massive galaxies that are selected as star forming via the $UVJ$ method. The black ``X" shows the median $U-V$ colour for the passive population and the black plus sign shows the median $U-V$ colour for the star forming population. The greyscale shows the whole UDS galaxy sample within each redshift bin. The red dashed line shows the 95\% stellar mass completeness limit, such that the sample
in each plot is complete to the right of this line.}
\label{UVmasshigh}  
\end{figure*}

\begin{figure*} 
\includegraphics[scale=0.65]{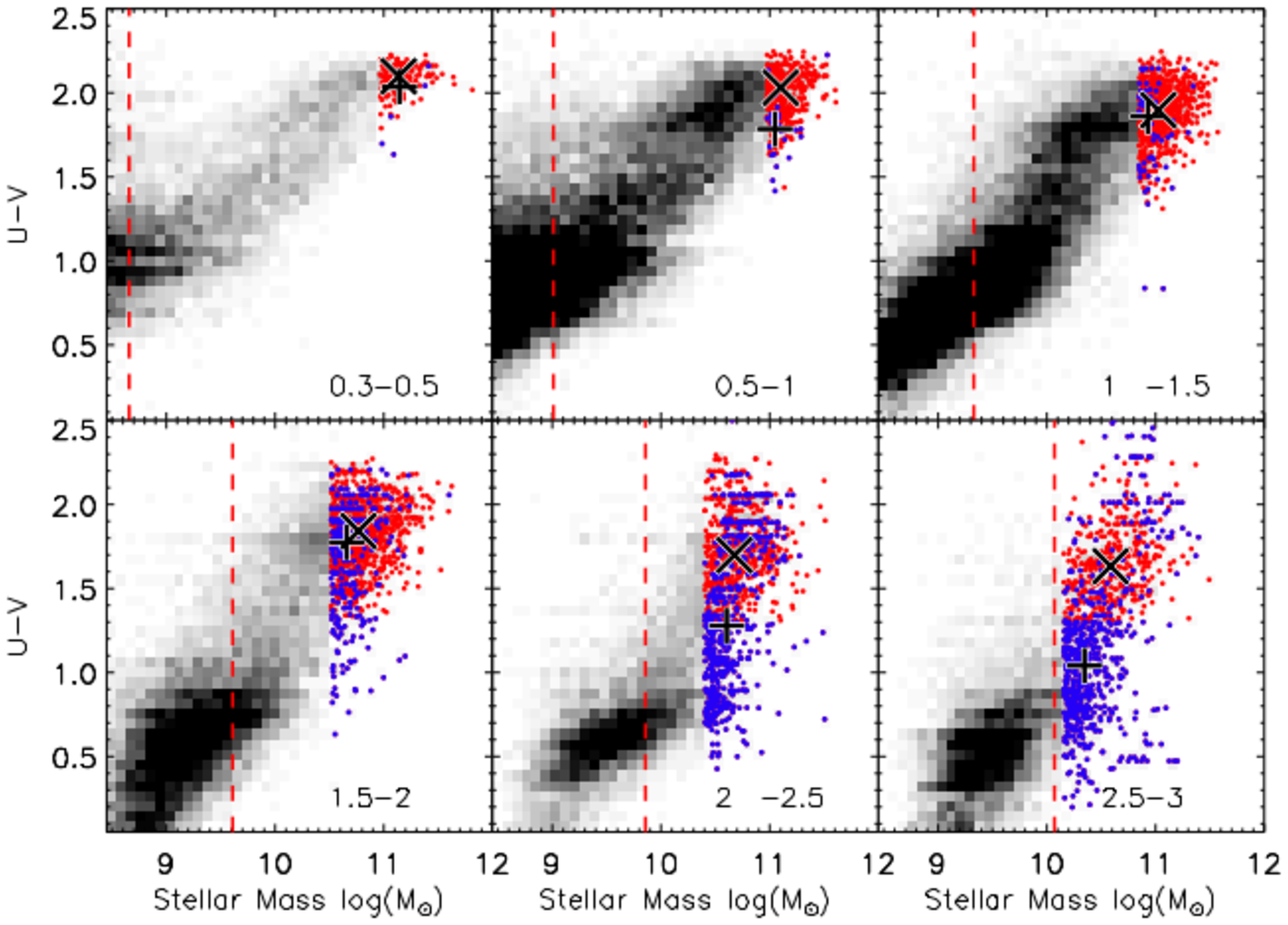}           
\caption[Stellar mass versus rest frame $U-V$ colour for all galaxies selected via the C\--GaND]{Stellar mass versus rest frame $U-V$ colour for all galaxies selected via the constant number density selected sample with $n=3\times10^{-4}\rm{Mpc^{-3}}$. The red circles show the progenitors of massive galaxies that are selected as passive via the $UVJ$ method. The blue circles show the progenitors of massive galaxies that are selected as star forming via the $UVJ$ method. The black ``X" shows the median $U-V$ colour for the passive population and the black plus sign shows the median $U-V$ colour for the star forming population. The greyscale shows the whole UDS galaxy sample within each redshift bin. The red dashed line shows the 95\% stellar mass completeness limit.}
\label{UVmasslow} 
\end{figure*}

\label{lastpage}
\end{document}